\documentclass[twocolumn,10pt]{asme2ej}

\usepackage{graphicx} %% for loading jpg figures
\usepackage{todonotes}
\usepackage[abs]{overpic}
\usepackage{subfig}
\usepackage{hyperref}

\usepackage{amssymb}
\usepackage{amsmath}

\newcommand{\ve}[1]{\text{$\mathbf#1$}} %vector
\newcommand{\D}[1]{\mathrm{d}{#1}}
\newcommand{\DSS}[1]{\mathrm{d}^2{#1}}
\newcommand{\DS}[1]{\mathrm{d}{#1}^2}

\title{{\fontfamily{phv} \fontsize{24pt}{24pt} \selectfont \textbf{The effect of compressibility on the behaviour of filter media}}}

\author{Jakub K{\"o}ry, Armin U.~Krupp, Colin P.~Please \& Ian M.~Griffiths\thanks{Address all correspondence to this author.}\\
    \affiliation{ 
    Mathematical Institute\\
	University of Oxford\\
	Oxford, OX2 6GG, UK\\
	Email: ian.griffiths@maths.ox.ac.uk
    }
}

\begin{document}

\maketitle    

%%%%%%%%%%%%%%%%%%%%%%%%%%%%%%%%%%%%%%%%%%%%%%%%%%%%%%%%%%%%%%%%%%%%%%
\begin{abstract}
{\it
A filter comprises porous material that traps contaminants when fluid passes through under an applied pressure difference. One side-effect of this applied pressure, however, is that it compresses the filter. This changes the permeability, which may affect its performance. As the applied pressure increases the flux of fluid processed by the filter will also increase but the permeability will decrease. Eventually the permeability reaches zero at a point in the filter and the fluid flux falls to zero. In this paper we derive a model for the fluid transport through a filter due to an applied pressure difference and the resulting compression. We use this to determine the maximum operating flux that can be achieved without the permeability reaching zero and the filter shutting down. We determine the material properties that balance the desire to maximize flux while minimizing power use. We also show how choosing an initial spatially dependent permeability can lead to a uniformly permeable filter under operation and we find the permeability distribution that maximizes the flux for a given applied pressure, both of which have desirable industrial implications. The ideas laid out in this paper set a framework for modelling more complex scenarios such as filter blocking. %\note{139/250 words.}

}
\end{abstract}

%%%%%%%%%%%%%%%%%%%%%%%%%%%%%%%%%%%%%%%%%%%%%%%%%%%%%%%%%%%%%%%%%%%%%%
%\begin{nomenclature}
%\entry{A}{You may include nomenclature here.}
%\entry{$\alpha$}{There are two arguments for each entry of the nomemclature environment, the symbol and the definition.}
%\note{To do}
%\end{nomenclature}

%%%%%%%%%%%%%%%%%%%%%%%%%%%%%%%%%%%%%%%%%%%%%%%%%%%%%%%%%%%%%%%%%%%%%%
\section{Introduction}

Dead-end filtration is a process by which contaminants from a fluid are removed by a filter whose surface is perpendicular to the flow direction. Common applications in industry range from the small-scale, such as protein filtration and virus removal, to large-scale processes such as wastewater treatment~\cite{Mulder_BasicPrinciplesOfMembraneTechnology}.
Rainwater permeating through granular beds is an instance of dead-end filtration observed in nature. Contaminants may be removed either by steric effects (trapping of particles) or by adsorption onto the surface. Porous materials represent excellent filters since they are both able to trap contaminants in the void space and have a very large surface-area-to-volume ratio for adsorption~\cite{Bear1972,wu2012experimental}.  

For a fluid to flow through a porous medium, gradients in pressure must exist, exerting a net force on the porous matrix which deforms as a result. The earliest attempts to model the mechanics of deformable, fluid-filled, porous media date back to the studies of one-dimensional (laterally constrained) soil consolidation by Terzaghi~\cite{Terzaghi1925}. These studies subsequently led to the establishment of the standard equations of poroelasticity by Biot, who first generalized the analysis to three dimensions~\cite{Biot1941} and later to anisotropic materials~\cite{Biot1955}. In the following decades, the underlying theory has been further developed using theory of solid--fluid mixtures~\cite{Kenyon1976a,Kenyon1976b} and also rederived via the method of homogenization~\cite{Burridge1981}. Since then, poroelasticity has been applied not only in hydrogeology but in a range of other applications, such as biological~\cite{Kenyon1979} and filtration problems~\cite{Parker1987}. 

The interaction between the deformation of a porous matrix and the fluid flow in such a situation is fully coupled: the flowing fluid deforms the material and the deformation of the porous matrix in turn affects its permeability and thus the fluid flow. (The latter relationship, however, has received considerably less attention in the scientific literature.) To our knowledge, the first attempt to include the effects of deformation on flow through a porous medium was presented in~\cite{Parker1987}, in which the analytical solutions to poroelasticity equations were found under three different constitutive assumptions, namely that the local permeability was considered to be a constant, linear or exponential function of the local strain. The authors also studied the case where the effective stiffness of the porous matrix depends non-trivially on the applied strain.

Poroelasticity equations remain an important modelling approach to describe and explain experimentally observed interactions between fluids and solids on the macroscale. A combination of experimental measurements and poroelasticity theory was used in \cite{MacMinn2015} to study the deformation effects that a fluid has on a dense monolayer of soft particles into which it is injected. In \cite{Hewitt2016}, experiments with water flowing through a porous medium composed of small, hydrogel spheres are used to test a two-phase model of flow and compression. In  \cite{Herterich2019}, a model was developed to describe the implications of elastic deformation in an annular filter. The model showed how the filter pores stretched during operation, enabling them to trap larger particles than expected. On reversing the flow to clean the filter (backflushing) the pores relaxed back to their undeformed size, holding the larger particles in place and preventing them from being removed. This was used to explain experimentally observed filter blocking behaviour that was not reversed by periodic backwashing.

%\note{Ian: 1. In this paper, Hewitt mentions regarding poro-elasticity "...and continuing interest in the subject has, in part, been spurred by industrial applications to enhanced oil recovery, carbon dioxide sequestration and hydraulic fracturing", but doesn't cite any sources. Should we mention these applications? 2. Moreover, could you write 1-2 sentences about \cite{Herterich2019} (as you were one of the authors for this one)? This could become a bridge when we go back from poroelasticity to filtration applications in the next paragraph maybe? 3. Could you maybe also double-check I haven't missed out anything important in recent progress section? To me, it seems like Duncan Hewitt did lots of kind-of-related stuff, but most often with high Rayleigh number (turbulence?); convection-diffusion on top of Darcy. So I don't think we need to/should cite these.}

Despite considerable progress and the broad applications of poroelasticity theory, analytic solutions providing deeper insight into such problems are rare. Furthermore, the implications of the flow-induced deformation on filtration efficiency have not yet been fully understood, especially in the case where the rest-state permeability of the filter (under no applied pressure difference) is non-uniform. The main purpose of this work is to achieve a greater understanding of the implications of filter compressibility through analytical solutions of an idealized filter in a practical scenario. 

The paper is organized as follows. In Section~\ref{sec:filter_uniform} we present the equations governing the fluid flow and the elastic response of the porous medium. In Section~\ref{Section:Solution} we solve these for a porous medium whose permeability is initially uniform, to determine the distribution of compression in the medium when a pressure difference is applied to drive fluid through it. A similar problem is modelled and experimentally studied in \cite{Parker1987} and \cite{Hewitt2016}. However, here our focus is on  how the compressibility of the material affects the filter performance and the impact of shutdown of the filter when compressed beyond a critical value. In Section~\ref{subsec:optimization_no_cake_uniform_starting_perm} we consider the optimal performance of the filter when balancing the desire to maximize flux and minimize power use. 
In Section~\ref{sec:inhomo} we relax the constraint of an initially uniform permeability with the view of determining the initial rest-state permeability that leads to uniform permeability under operation (Section~\ref{subsub:uniform_post}) and the initial permeability gradient that maximizes the flux under operation (Section~\ref{A general linear rest-state permeability}). We consider two types of material of interest: one in which the permeability changes proportionally to the rest-state permeability when compressed and a second whose permeability varies independently of the rest-state permeability. In both cases we find that we cannot achieve a uniform operating permeability while maximizing the flux and so face a choice in which of these is most favourable in a particular scenario.

We conclude in Section~\ref{Section:Conclusions} and discuss implications of the work and how this can be used as a framework for tackling the behaviour of compressible filters as they trap contaminants.

\section{Modelling the compression of a porous medium by fluid flow}
\label{sec:filter_uniform}
We begin by deriving the governing equations for the deformation of the porous medium and propose a constitutive relation between the deformation and the permeability of the medium.
%We first analyse the equations in the case where the permeability is unaffected by the deformation and then when it depends on the deformation.

\subsection{Governing equations}
\label{Section:Governing equations}

 Neglecting gravity and assuming that the deformations are small compared with the medium size, the standard linear theory of poroelasticity (for details see  \cite{Fowler1997} or \cite{Howell2009}) gives the Navier equation for deformation of porous media due to fluid flow
\begin{equation}\label{eq:Navier-equilibrium equation}
 (\tilde{\lambda}_{\mathrm{eff}} + \tilde{\mu}_{\mathrm{eff}}) \tilde{\nabla} (\tilde{\nabla} \cdot \tilde{\ve{u}}) + \tilde{\mu}_{\mathrm{eff}}\tilde{\nabla}^2 \tilde{\ve{u}} = \tilde{\rho} \frac{\partial^2 \tilde{\ve{u}}}{\partial \tilde{t}^2} + \tilde{\nabla} \tilde{p},
\end{equation}
where $\tilde{\ve{u}}$ denotes the displacement of the porous medium from its equilibrium, $\tilde{p}$ denotes the pressure in the fluid, $\tilde{\rho}$, $\tilde{\lambda}_{\mathrm{eff}}$ and $\tilde{\mu}_{\mathrm{eff}}$ denote the density and the effective elastic constants of the material respectively and $\tilde{t}$ denotes time. Note that tildes indicate the use of dimensional variables throughout this paper. The pressure-gradient term represents the drag exerted by the fluid on the solid matrix. Here, we assume that $\tilde{\rho}$, $\tilde{\lambda}_{\mathrm{eff}}$ and $\tilde{\mu}_{\mathrm{eff}}$ are all constant.

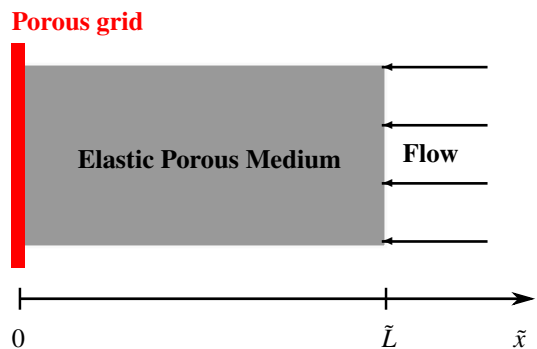
\begin{figure}
\vspace{8.5mm}
\centering
\begin{overpic}[width=0.40\textwidth,tics=10]{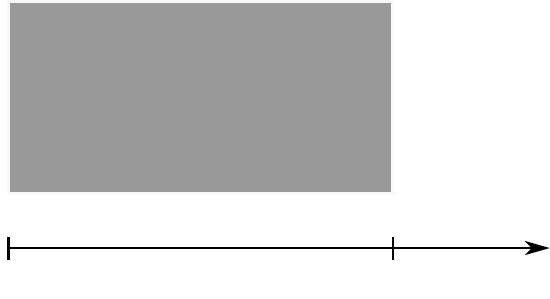}
\linethickness{1pt}
\normalsize{
\put(190,0){$\tilde x$}
\put(0,0){$0$}
\put(140,0){$\tilde{L}$}
\put(180,40){\vector(-1,0){40}}
\put(180,62){\vector(-1,0){40}}
\put(180,84){\vector(-1,0){40}}
\put(180,106){\vector(-1,0){40}}
\put(148,70){\textbf{Flow}}
\put(25,68){\textbf{Elastic Porous Medium}}
\put(0,120){\color{red}{\textbf{Porous grid}}}
\linethickness{5pt}
\color{red}
\put(0,30){\vector(0,1){85}}
}
\end{overpic}
\caption{Simplified model of the experiment.}
\label{fig:CompressionModel}
\end{figure}

We consider the idealized case of a medium  attached to a porous grid at $\tilde{x} = 0$, through which a fluid flows uniformly (see Figure \ref{fig:CompressionModel}). The grid is assumed to offer no resistance to the flow. 
Assuming that the poroelastic response time within the (one-dimensional) medium is much smaller than the operating time, we can neglect the transient behaviour (for small times after switching on the fluid flow), and thus we study a steady-state reduction of \eqref{eq:Navier-equilibrium equation}, i.e.
\begin{equation}\label{eq:OneDimEffectiveNavier}
(\tilde{\lambda}_{\mathrm{eff}} + 2\tilde{\mu}_{\mathrm{eff}}) \frac{\mathrm{d}^2 \tilde u}{\mathrm{d} \tilde{x}^2} = \frac{\mathrm{d} \tilde{p}}{\mathrm{d} \tilde{x}},
\end{equation}
where $\tilde{u}$ and $\tilde{p}$ are defined on the undeformed domain $\tilde{x} \in (0,\tilde{L})$ and $\tilde{L}$ denotes the undeformed size of the medium. The fluid flux $\tilde{q}$ (volumetric flow rate per unit surface area) through the porous medium is determined by Darcy's law,
\begin{equation}\label{eq:Compression_DarcyLaw}
\tilde q = \frac{\tilde{k}}{\tilde{\eta}} \frac{\mathrm{d} \tilde p}{\mathrm{d}\tilde x},
\end{equation}
where the viscosity $\tilde{\eta}$ of the fluid is assumed to be constant. We assume the fluid to be incompressible and therefore conservation of mass reads
\begin{equation}\label{eq:Compression_IncompressibleFluid}
\frac{\mathrm{d} \tilde q}{\mathrm{d} \tilde x} = 0,
\end{equation}
and so the flux is constant. We allow the permeability $\tilde k$ to depend on the deformation of the medium $\mathrm{d} \tilde u / \mathrm{d} \tilde x$. While there does not exist one general constitutive relation between the permeability and the porosity of a porous medium, the two parameters are usually closely related. 
Here we propose a linear constitutive relationship 
\begin{equation}\label{eq:ConstitutiveRelationK}
\tilde k \left(\frac{\mathrm{d} \tilde u}{\mathrm{d} \tilde x}\right) = \tilde{k}_1 + \tilde{k}_2 \frac{\mathrm{d} \tilde u}{\mathrm{d} \tilde x},
\end{equation}
which will hold for small deformations. Here, $\tilde{k}_1$ denotes the permeability of the medium in its undeformed state, while a local deformation $\mathrm{d} \tilde u / \mathrm{d} \tilde x$ changes the permeability by a factor $\tilde{k}_2$ (which has units of permeability).
While $\tilde{k}_1$ is positive, $\tilde{k}_2$ can in principle be positive, negative or zero but here we will restrict our analysis to the cases where $\tilde{k}_2 \ge 0$. Note that the same constitutive relationship was used in \cite{Parker1987} (see Equation (3.11) therein).  We will begin by assuming that $\tilde{k}_1$ and $\tilde{k}_2$ are both constant. We will relax this assumption in Section~\ref{sec:inhomo} to allow $\tilde{k}_1$ to vary in space so that we can cater for filters with non-uniform initial permeabilities, and for $\tilde{k}_2$ to vary in space so that the way in which the material permeability changes can depend on the underlying structure.

Since $\D{\tilde{u}}/\D{\tilde{x}} \geq -1$ by definition of $\tilde{u}$, the permeability in \eqref{eq:ConstitutiveRelationK} is able to reach zero for an admissible strain if
\begin{equation}
\label{eqn:shutdown_constraint}
\tilde{k}_2 \geq \tilde{k}_1.
\end{equation}
Since the medium is attached to the porous grid at $ \tilde x = 0$, the displacement at this point must be zero.
The right end $\tilde x = \tilde{L}$ is free and therefore cannot be compressed, leading to the two boundary conditions
\begin{equation}\label{eq:BCDisplacement}
\tilde u(0) = 0 \hspace{0.1cm} \text{(no displacement)} \hspace{0.1cm} \text{and} \hspace{0.1cm} \frac{\mathrm{d} \tilde u}{\mathrm{d} \tilde x} (\tilde{L}) = 0 \hspace{0.1cm} \text{(free end)}.
\end{equation}
We prescribe a fixed pressure on the left- and right-hand edges of the medium:
\begin{equation}
\tilde p(0) = \tilde{p}_{\mathrm{out}} \quad \text{and}\quad  \tilde p(\tilde{L}) = \tilde{p}_{\mathrm{in}}.
\end{equation}
When $\tilde{p}_{\mathrm{in}}>\tilde{p}_{\mathrm{out}}$ the medium is under compression and the strain field $\mathrm{d} \tilde u / \mathrm{d} \tilde x \leq 0$ and $\tilde{u} \leq 0$ while the flux $\tilde q \geq 0$ $\forall \tilde{x} \in [0,\tilde{L}]$. We note that our model also holds when  $\tilde{p}_{\mathrm{in}}<\tilde{p}_{\mathrm{out}}$, in which case the filter will be stretched out, but this is of less interest in this paper.  

%As the filter is held fixed at $\tilde{x} = 0$ and we expect the largest (in absolute value) displacement to occur at the free end $\tilde{x} = \tilde{L}$, the small-deformations assumption, needed for our linear poroelastic model \eqref{eq:Navier-equilibrium equation} to be valid, reads
%\begin{equation}\label{small_deform_dim}
%\frac{-\tilde{u}(\tilde{x}=L) {\tilde{L}} \ll 1.
%\end{equation}

\subsection{Nondimensionalization}

We nondimensionalize the problem by setting
\begin{equation}
\label{eq:nondimensionalization 1}
\tilde{x} = \tilde{L} x, \quad \tilde{p} = \left(\tilde{p}_{\mathrm{in}} - \tilde{p}_{\mathrm{out}}\right)p+\tilde{p}_{\mathrm{out}}, \quad \tilde{u} = \frac{\tilde{p}_{\mathrm{in}} - \tilde{p}_{\mathrm{out}}}{\tilde{\lambda}_{\mathrm{eff}} + 2\tilde{\mu}_{\mathrm{eff}}} \tilde{L} u,
\end{equation}
and
\begin{equation}\label{eq:NonDimPermAndFlow}
\tilde{q} = \frac{\tilde{k}_1 \left(\tilde{p}_{\mathrm{in}}, - \tilde{p}_{\mathrm{out}}\right)}{\tilde{\eta} \tilde{L}}q \qquad \quad \tilde{k}= \tilde{k}_1 k.
\end{equation}
Equation \eqref{eq:OneDimEffectiveNavier} then becomes
\begin{equation}\label{eq:NonDimNavierEquation}
\frac{\DSS{u}}{\DS{x}} = \frac{\D{p}}{\D{x}},
\end{equation}
and Darcy's law \eqref{eq:Compression_DarcyLaw} reads
\begin{equation}
\label{eqn:Darcy_rescaled}
 \bigg( 1 + \gamma \frac{\D{u}}{\D{x}} \bigg) \frac{\D{p}}{\D{x}} = q.
\end{equation}
Here, the flux $q$ is constant due to \eqref{eq:Compression_IncompressibleFluid} and 
\begin{equation}\label{def_of_gamma}
\gamma = \kappa \Delta p \geq 0,
\end{equation}
where 
\begin{align}\label{def_kappa_deltap}
    \kappa = \frac{\tilde{k}_2}{\tilde{k}_1}, \quad \qquad \Delta p = \frac{\tilde{p}_{\mathrm{in}} - \tilde{p}_{\mathrm{out}}}{\tilde{\lambda}_{\mathrm{eff}} + 2\tilde{\mu}_{\mathrm{eff}}}.
\end{align}
The dimensionless boundary conditions are given by
{\refstepcounter{equation}
\[
\label{eq:Compression_Rescaled_BC}
 u(0) = 0,  \quad \frac{\mathrm{d}u}{\mathrm{d}x}(1) = 0, \quad p(0) = 0,  \quad p(1) = 1. 
\eqno{(\theequation\textrm{a--d})}
\]}
With this nondimensionalization, the system behaviour is governed by a single dimensionless parameter, $\gamma$. This parameter can be interpreted as describing either: the material properties, \emph{i.e.}, increasing $\gamma$ corresponds to increasing the material's susceptibility to permeability change upon compression; or the operating conditions, \emph{i.e.}, increasing $\gamma$ corresponds to increasing the applied pressure across the filter. %Later on, we will use the decomposition $\gamma = \kappa \Delta p$ to separate the material and operating-condition effects.

\subsection{Restrictions}

We first state the restrictions on model parameters required so that the linear poroelasticity assumption underlying Equation \eqref{eq:Navier-equilibrium equation} is valid and so that one avoids negative permeabilities.

\subsubsection{The small-deformations assumptions}

Integrating Equation \eqref{eq:NonDimNavierEquation} over $0 \leq x \leq 1$ and using the boundary conditions (\ref{eq:Compression_Rescaled_BC}b--d), we obtain
\begin{equation}
\label{eq:strain_at_fixed_end_rescaled}
- \frac{\D{ u}}{\D{x}}(0) = 1,
\end{equation}
which gives the compression at the fixed end $x=0$ and indeed the maximum compression across the whole medium. Using (\ref{eq:Compression_Rescaled_BC}a) and \eqref{eq:strain_at_fixed_end_rescaled} gives
\begin{equation}
\label{eq:u(1) requirement}
u(1) =  \int\limits_0^1 \frac{\D{u}}{\D{x}} \D{x} \geq -1.
\end{equation}
In dimensionless terms, the small-deformations approximation requires  $\vert u(1)\vert \ll 1/\Delta p$, which, combining with \eqref{eq:u(1) requirement}, is satisfied provided we assume $0< \Delta p \ll 1$.

%With this assumption, we expect $u,q,p = O(1)$.  
%Note that this restriction has no implications on $\gamma$; in this paper we are primarily interested in materials for which even small deformations can lead to shutdown of the medium (that is, $k=0$ at some position in the medium), which can only happen if $\kappa \gg 1$ so that $\gamma = O(1)$. (In Appendix B we show that this restriction requires a low material porosity.) 

\subsubsection{Avoiding negative permeabilities}

Since the permeability must always be non-negative, Equation \eqref{eq:ConstitutiveRelationK} provides the constraint
\begin{equation}\label{eq:Physical_BoundU_tilde}
\frac{\mathrm{d}u}{\mathrm{d}x} \geq -\frac{1}{\gamma}.
\end{equation} 
The maximum compression is expected at the grid $x=0$ with the value obtained in \eqref{eq:strain_at_fixed_end_rescaled}, which means that our model is only applicable when $\gamma \leq 1$ or, in dimensional terms,
\begin{equation}\label{p_constraint_shutdown}
\tilde{p}_{\mathrm{in}} - \tilde{p}_{\mathrm{out}} \leq \frac{\tilde{k}_1}{\tilde{k}_2}\left(\tilde{\lambda}_{\mathrm{eff}} + 2 \tilde{\mu}_{\mathrm{eff}}\right).
\end{equation}
When equality is attained, the permeability of the medium at $x=0$ reaches zero.
Beyond this threshold, the fluid flow ceases. Thus, choosing $\gamma = 1$ corresponds to the maximum possible pressure difference across the medium for which there exists a fluid flow, and we will find that the flux at this value is maximal. Note that this is consistent with the result from \cite{Parker1987}. However, they continue their analysis for $\gamma >1$ and also erroneously obtain flow in the regime when the porous medium would in principle have shut down (see Appendix A for more details).
%, as $\gamma$ here corresponds to $1/\hat{R}$ in \cite{Parker1987} and with our nondimensionalization, we have that $\tilde{p}_1$ from \cite{Parker1987} equals $1$.

\section{Solution}
\label{Section:Solution}

\begin{figure}
\centering
\subfloat[Strain of the medium]{\begin{overpic}[width=0.24\textwidth,tics=10]{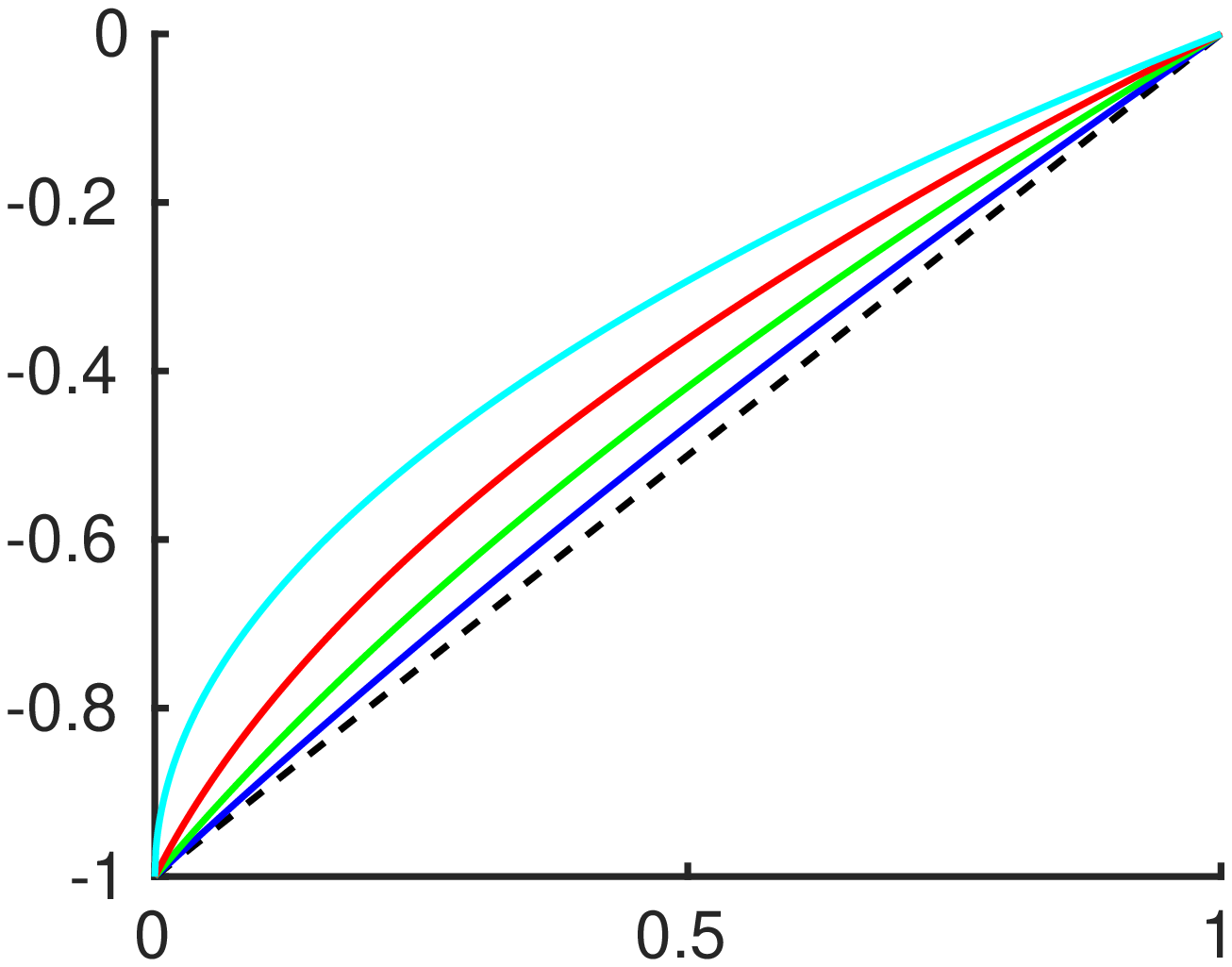}
\put(95,0){$x$}
\put(10,92){$\frac{\D{u}}{\D{x}}$}
\put(60,35){\vector(-1,1){25}}
\put(18,68){Increasing $\gamma$}
\end{overpic}}
\subfloat[Displacement of the medium]{\begin{overpic}[width=0.24\textwidth,tics=10]{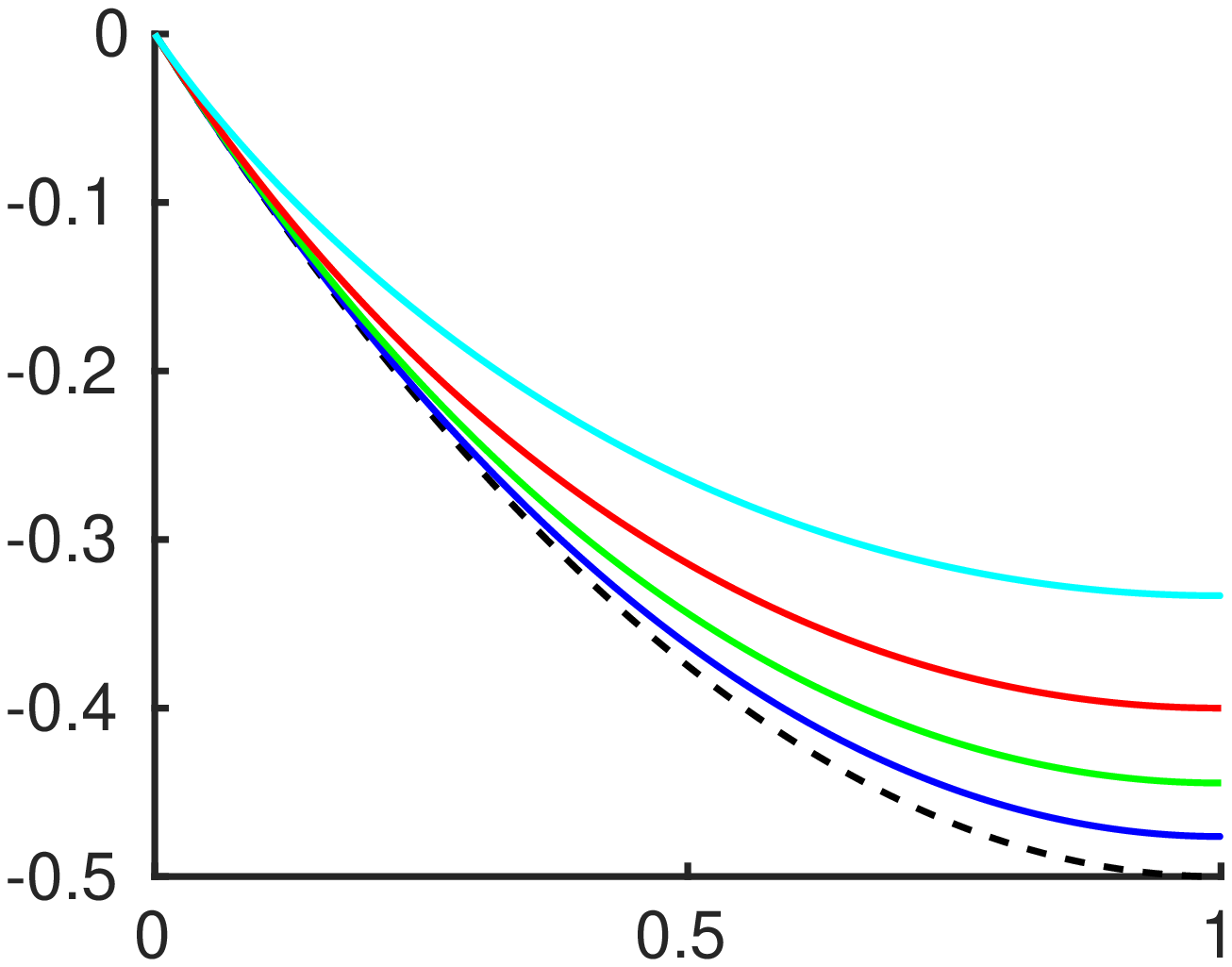}
\put(95,0){$x$}
\put(11,92){$u$}
\put(53,23){\vector(1,1){25}}
\put(60,57){Increasing $\gamma$}
\end{overpic}}
\caption{Displacement and strain of the medium with compression-dependent permeability for $\gamma = 0.25\ (\mathrm{blue}),\ 0.5\ (\mathrm{green}),\ 0.75\ (\mathrm{red}),\ 1\ (\mathrm{cyan})$. The incompressible ($\gamma=0$) case is shown for comparison in dashed black.}
\label{fig:CompressionOffilter}
\end{figure}

Since we are concerned with compressible porous media, we will assume $\gamma \neq 0$ (the case when $\gamma = 0$ is straightforward to solve). Substituting \eqref{eq:NonDimNavierEquation} into \eqref{eqn:Darcy_rescaled} gives
\begin{equation}\label{eqn:Darcy_rescaled_subbed}
\frac{\DSS{u}}{\DS{x}}\left(1 + \gamma \frac{\D{u}}{\D{x}}\right)  = \frac{\DSS{u}}{\DS{x}} + \frac{\gamma}{2} \frac{\mathrm{d}}{\D{x}} \bigg[\left(\frac{\D{u}}{\D{x}}\right)^2\bigg] = q,
\end{equation}
which when integrated yields
\begin{equation}\label{eqn:Navier_integrated}
\frac{\gamma}{2}\left(\frac{\D{u}}{\D{x}}\right)^2 + \frac{\D{u}}{\D{x}} = (x-1)q,
\end{equation}
where we have applied boundary condition (\ref{eq:Compression_Rescaled_BC}b). Together with Darcy's law \eqref{eqn:Darcy_rescaled} and equipped with boundary conditions \eqref{eq:Compression_Rescaled_BC}, this problem can be solved directly for $u$, $p$ and $q$ and we conclude that the resulting flux is
\begin{equation}\label{eq:Compression_NonDim_QDependentOnP0}
q = 1-\frac{\gamma}{2},
\end{equation}
the strain is
\begin{equation}\label{eq:ClosedFormOfPTuPTx_2}
\frac{\D{u}}{\D{x}} =  -\frac{1}{\gamma} \bigg( 1 - \sqrt{1+\gamma (\gamma -2) (1-x)} \bigg),
\end{equation}
the deformation is
\begin{align}
u(x) = - \frac{x}{\gamma} + \frac{2 (1 + \gamma(\gamma-2))^{3/2} - 2 (1 + \gamma (\gamma-2) (1-x))^{3/2} }{3 \gamma^2 (\gamma-2)},
\end{align}
using (\ref{eq:Compression_Rescaled_BC}a) and \eqref{eq:ClosedFormOfPTuPTx_2}, and the pressure is
\begin{align}
p(x) = 1 -\frac{1}{\gamma} \bigg( 1 - \sqrt{1+\gamma (\gamma -2) (1-x)} \bigg)
\end{align}
using \eqref{eqn:Darcy_rescaled} and (\ref{eq:Compression_Rescaled_BC}c). From \eqref{eq:Compression_NonDim_QDependentOnP0}, we observe that the maximum (dimensionless) flux $q$ is attained in the limit of incompressible media, $\gamma \to 0$. Moreover, Figure~\ref{fig:CompressionOffilter} suggests that $\gamma=0$ also results in the maximum compression $- \D{u}/\D{x}$ and absolute value of the displacement $\lvert u \rvert$ everywhere in the medium. This is sensible, since decreasing $\gamma$ should be interpreted as decreasing $\kappa$, and thus the sensitivity of the medium permeability to its deformation. %In order to study the effects of varying the pressure difference, $\Delta p$, while keeping $\kappa$ constant, in the next section we use the decomposition $\gamma= \kappa \Delta p$.

\section{Optimizing filter performance}
\label{subsec:optimization_no_cake_uniform_starting_perm}

\begin{figure}
\centering
\subfloat[Flux through the filter and power required for varying $\gamma$]{\begin{overpic}[width=0.24\textwidth,tics=10]{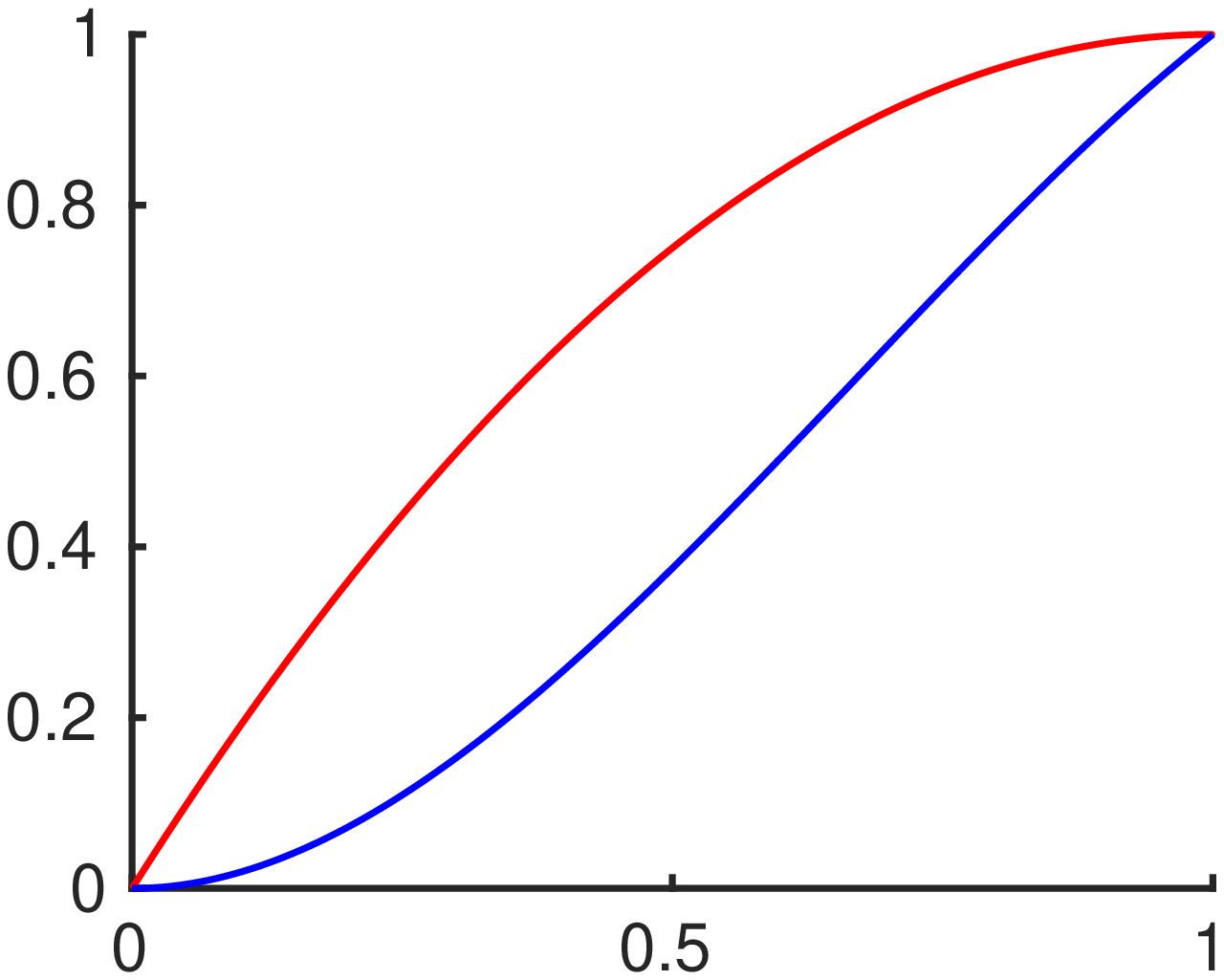}
\put(95,0){$\gamma$}
\put(37,54){$\mathcal{Q}$}
\put(65,30){$\mathcal{W}$}
\end{overpic}}
\subfloat[Power $\mathcal{W}$ required to generate a given flux $\mathcal{Q}$]{\begin{overpic}[width=0.24\textwidth,tics=10]{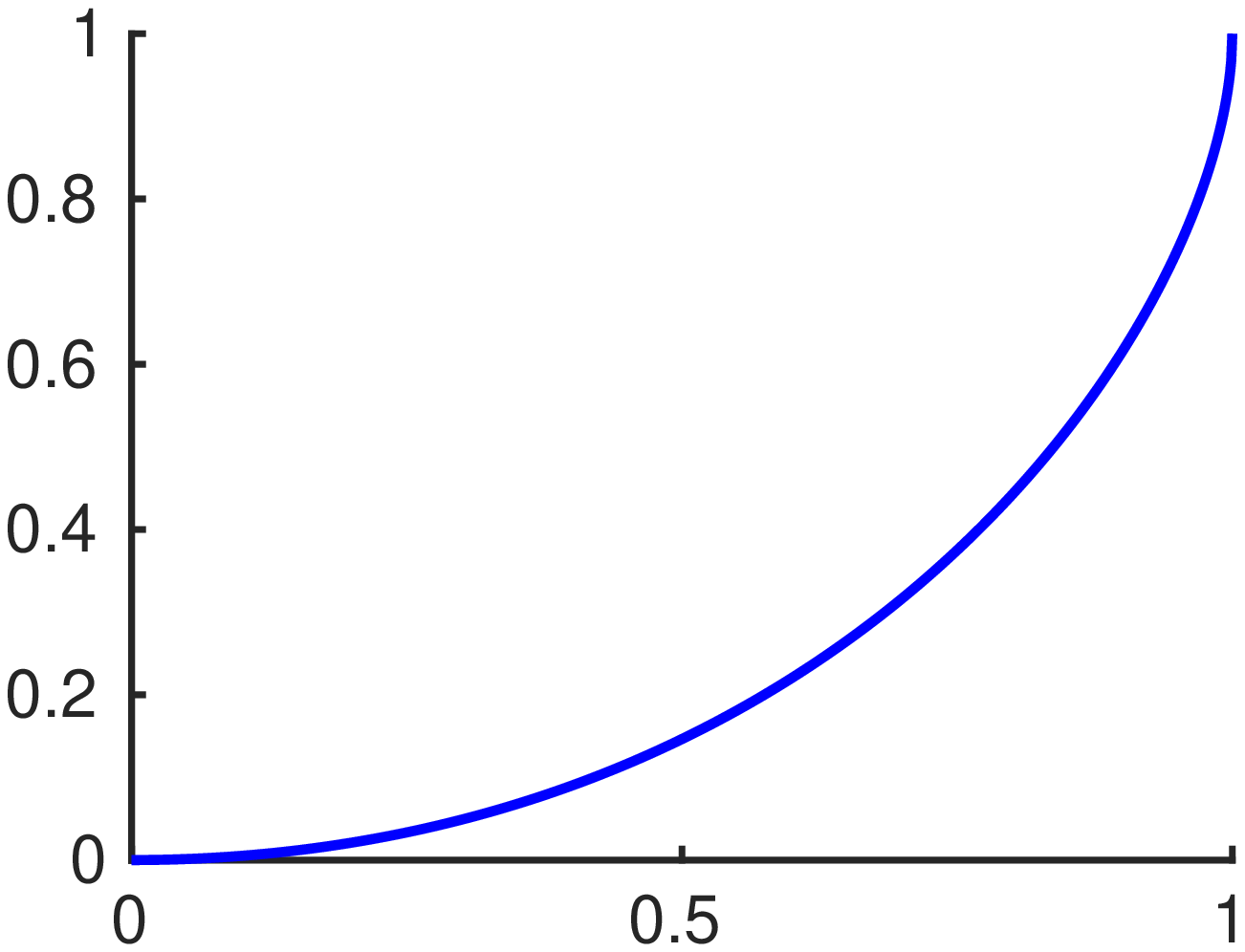}
\put(95,0){$\mathcal{Q}$}
\put(10,85){$\mathcal{W}$}
\end{overpic}}\\
\subfloat[$\gamma_{\mathrm{opt}}$ as a function of $\xi$]{\begin{overpic}[width=0.24\textwidth,tics=10]{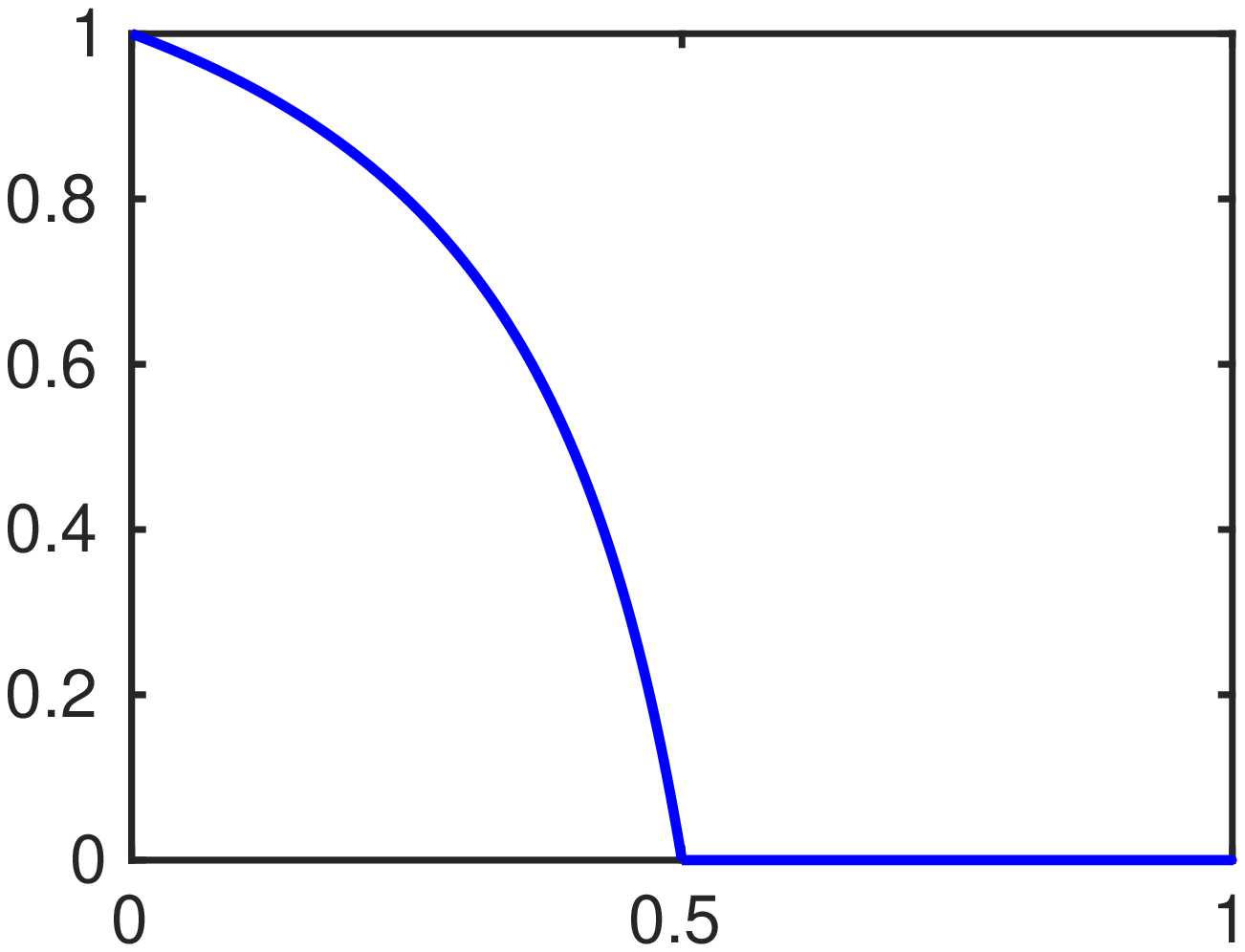}
\put(95,0){$\xi$}
\put(12,85){$\gamma_{\mathrm{opt}}$}
\end{overpic}}
\caption{(a) Both the flux $\mathcal{Q}$ and the power expended $\mathcal{W}$ are increasing functions of $\gamma$. (b) Variation of the optimum value of $\gamma$ as given by \eqref{eqn:gamma_opt} using objective functions $\mathcal{F}_{\xi}$ defined in \eqref{F_alpha_gamma_opt}. (c) Higher powers $\mathcal{W}$ lead to higher fluxes $\mathcal{Q}$, as given by \eqref{eqn:W_vs_Q}.}
\label{fig:TQofY}
\end{figure}

Having established the governing equations and physical limits we are now in a position to understand how the filter compressibility may be best used to provide optimal performance. To achieve this, we consider the relationship between the pressure applied across the porous medium and the resulting flux and power required to drive the flow. We first note that the dimensionless 
flux $q$ defined in \eqref{eq:NonDimPermAndFlow} contains the pressure difference $\tilde{p}_{\mathrm{in}}-\tilde{p}_{\mathrm{out}}$. While 
this choice of nondimensionalization led to a system that may be described in terms of a single parameter, $\gamma$, in order to establish the flux--pressure difference relationship it is necessary to introduce a new flux that is not scaled with the pressure difference. To this end we define a new dimensionless flux  $\mathcal{Q}$ via
\begin{equation}\label{new_dimless_flux}
\tilde{q} =\frac{\tilde{k}_1^2 (\tilde{\lambda}_{\mathrm{eff}} + 2 \tilde{\mu}_{\mathrm{eff}})}{2 \tilde{\eta} \tilde{L} \tilde{k}_2} \mathcal{Q}.
\end{equation}
Changes in the dimensional flux due to variations in the pressure difference (without changing any material properties) will then be appropriately expressed through the dimensionless measure of flux $\mathcal{Q}$. Comparing the two dimensionless fluxes from \eqref{eq:NonDimPermAndFlow} and  \eqref{new_dimless_flux}, we get $\mathcal{Q} = 2 \gamma q$. Thus, \eqref{eq:Compression_NonDim_QDependentOnP0} implies
\begin{equation}\label{flow_rate_dimless_new}
\mathcal{Q}= \gamma (2-\gamma)
\end{equation}
and so $\mathcal{Q}$ is maximized at the threshold value $\gamma = 1$. %From Equation \eqref{p_constraint_shutdown} we see that the maximum applicable pressure difference only depends on the material parameters and not on the thickness of the filter.
%The maximum flux (or flow rate), however, does depend on the thickness of the filter. 
\\
One desirable industrial outcome is to minimize the power per unit area of membrane required to generate a certain flux,
\begin{equation}
\tilde{W}= \tilde{q} (\tilde{p}_{\mathrm{in}}-\tilde{p}_{\mathrm{out}}). % =  \frac{\tilde{k}_1^2 (\tilde{\lambda}_{\mathrm{eff}} + 2 \tilde{\mu}_{\mathrm{eff}})}{2 \tilde{\eta} \tilde{k}_2 \tilde{L}} \mathcal{Q} \Delta p (\tilde{\lambda}_{\mathrm{eff}} + 2 \tilde{\mu}_{\mathrm{eff}}) ,
\end{equation}
Nondimensionalizing via 
\begin{align}
    \tilde{W}=\frac{\tilde{k}_1^3 (\tilde{\lambda}_{\mathrm{eff}} + 2 \tilde{\mu}_{\mathrm{eff}})^2}{2 \tilde{\eta} \tilde{L} \tilde{k}_2^2}\mathcal{W}
\end{align}
we find 
\begin{align}
\label{power_dimless_new}
    \mathcal{W}=(2-\gamma)\gamma^2
\end{align} 
using \eqref{flow_rate_dimless_new} and thus, given $0 \le \gamma \le 1$, $\mathcal{W}$ attains its minimum when $\gamma=0$. 
Equations \eqref{flow_rate_dimless_new} and  \eqref{power_dimless_new} therefore imply that 
the flux is maximized when $\gamma$ is maximized but the power required is minimized when $\gamma$ is minimized (see Figure~\ref{fig:TQofY}a). 
Thus, we cannot find a single value of $\gamma$ that both maximizes the flux and minimizes the required power. However, we can obtain a relationship between $\mathcal{W}$ and $\mathcal{Q}$ by combining \eqref{flow_rate_dimless_new} with \eqref{power_dimless_new} and eliminating $\gamma$, which results in
\begin{equation}\label{eqn:W_vs_Q}
\mathcal{W} = \mathcal{Q} \left( 1 - \sqrt{1-\mathcal{Q}} \right)
\end{equation}
and the resulting functional dependence is plotted in Figure \ref{fig:TQofY}(b). To address the goal of making $\mathcal{Q}$ as large as possible while making $\mathcal{W}$ as small as possible one can introduce the following family of objective functions
\begin{equation}\label{F_alpha_gamma_opt}
\mathcal{F}_{\xi}(\gamma) = \frac{\mathcal{Q}}{\mathcal{W}^\xi} = \gamma^{1-2\xi} (2-\gamma)^{1-\xi}.
\end{equation}
Here, $\xi\in (0,1)$ provides a measure of the relative importance of power expended compared with the resulting flux through the filter, and will differ depending on the requirements of a given scenario. For every such $\xi$ we can find that $\gamma_{\mathrm{opt}}(\xi)$ maximizing $\mathcal{F}_{\xi}(\gamma)$ satisfies (see Appendix B for details)
\begin{equation}\label{eqn:gamma_opt}
    \gamma_{\mathrm{opt}}= \bigg\{\begin{array}{lr}
        \displaystyle\frac{2 (2 \xi -1)}{3 \xi -2} & \text{for } 0\leq \xi \leq 1/2, \\
        0 & \text{for } 1/2 \leq \xi \leq 1,
        \end{array}
\end{equation}
which  is plotted in Figure \ref{fig:TQofY}(c). We notice that
for $\xi \ge 1/2$, $\gamma_{\textrm{opt}}=0$, which corresponds to zero flux. This indicates that if we prize energy expenditure over flux beyond a certain critical point then the optimal operating regime is not to filter at all.  

As noted in Section~\ref{Section:Governing equations}, we could also reverse the sign of the pressure so that $\tilde{p}_\mathrm{in}<\tilde{p}_\mathrm{out}$. In this case the filter would be stretched out by the flow passing through and the higher the pressure the larger the flux. The maximum applicable pressure difference would then be determined by a threshold strain at which the filter tears rather than considerations regarding the permeabilities.

\section{Inhomogeneous starting permeabilities}
\label{sec:inhomo}

%\note{Ian: Could you say something here about how such filters can exist. Point to stacking different filters with different permeabilities together as a first step. Also talk about Leeds Nonwovens where they make this.}
In the previous section we have seen how a medium whose permeability is spatially uniform in the rest state, i.e.~when the pressure difference is zero, can deform under an applied  pressure difference and lead to a spatially varying permeability under operation. The resulting non-uniform permeability during operation may also lead to uneven contaminant capture. 
In this section, we therefore explore the advantages of manufacturing a filter with a spatially heterogeneous starting permeability, which are of considerable interest in industry (see, for example \cite{Barg2009,Vida2012} and other references in \cite{Dalwadi2015}).

We identify two relevant types of permeability relation of interest: one in which, under compression, the permeability varies proportionally to the rest-state permeability; and another in which it varies independently of the rest-state permeability. For each of these cases we determine the rest-state permeability required either to obtain a uniform permeability during operation (Section~\ref{subsub:uniform_post}) or to maximize the flux during operation for a given family of rest-state permeabilities (Section~\ref{A general linear rest-state permeability}). 

%We will find that these two rest-state permeabilities are different, and so one cannot both maximize the flux and have an operating permeability that is uniform in space.

\subsection{Dependence of flux on rest-state permeability}

We relax the assumption on an initial spatially uniform permeability by assuming 
\begin{align}
\label{eq:k img}
    \tilde{ k}= \tilde{k}_1(\tilde x) + {\tilde{k}_2(\tilde{x})} \frac{\D{\tilde u}}{\D{\tilde x}},
\end{align}
and define the average of the rest-state filter permeability as
\begin{equation}
\label{eq:bar k}
\langle \tilde{k} \rangle = \frac{1} {\tilde{L}} \int\limits_0^{\tilde{L}} \tilde{k}_1(\tilde x) \D{\tilde x}.
\end{equation}
We nondimensionalize the problem as in \eqref{eq:nondimensionalization 1} but replace \eqref{eq:NonDimPermAndFlow} with 
\begin{align}\label{eq:NonDimPermAndFlow_inhomogeneous}
\tilde{q} &= \frac{\langle \tilde{k} \rangle \left(\tilde{p}_{\mathrm{in}}, - \tilde{p}_{\mathrm{out}}\right)}{\tilde{\eta} \tilde{L}}q, & \tilde{k}&= \langle \tilde{k} \rangle k,
\end{align}
noting that this reduces to \eqref{eq:NonDimPermAndFlow} when $\tilde{k}_1$ {and $\tilde{k}_2$ are} constant. 
The Navier equation \eqref{eq:OneDimEffectiveNavier},  Darcy's law \eqref{eq:Compression_DarcyLaw} and \eqref{eq:k img} now read
\begin{align}\label{DarcyInhomogeneous}
q = k\frac{\D{p}}{\D{x}} = k\frac{\DSS{u}}{\DS{x}},\\
\label{eq:k img dimensionless}
k=\kappa_1(x)+\bar{\gamma}{(x)}\frac{\mathrm{d}u}{\mathrm{d}x},
\end{align}
where we have introduced $ \bar{\gamma}{(x)} := \Delta p \tilde{k}_2{(\tilde{L}x)}/\langle \tilde{k} \rangle$ (so $\bar{\gamma}{(x)}=\gamma$ when $\tilde{k}_1$ {and $\tilde{k}_2$ are} constant) and
\begin{equation}
\kappa_{1}(x) = \frac{\tilde{k}_1(\tilde{L} x)}{\langle \tilde{k} \rangle},
\end{equation}
for which we note that 
\begin{align}
\label{eq:mean permeability}
    \int\limits_0^1 \kappa_{1}(x) \,\mathrm{d}x = 1,
\end{align}
from \eqref{eq:bar k}. 

We must solve the system \eqref{DarcyInhomogeneous}--\eqref{eq:k img dimensionless} subject to the boundary conditions 
\eqref{eq:Compression_Rescaled_BC}, which we restate here for convenience, 
{\refstepcounter{equation}
\[
\label{BC: u,p img}
 u(0) = 0, \quad \frac{\mathrm{d}u}{\mathrm{d}x}(1) = 0, \quad p(0) = 0, \quad p(1) = 1. 
\eqno{(\theequation\textrm{a--d})}
\]}
Integration of \eqref{DarcyInhomogeneous} and using (\ref{BC: u,p img}b) and \eqref{eq:strain_at_fixed_end_rescaled} gives
\begin{equation}\label{target}
q =  \left[ \int\limits_0^1 \left( \kappa_1(x) + \bar{\gamma}{(x)} \frac{\D{u}}{\D{x}} \right)^{-1} \D{x} \right]^{-1}
\end{equation}
and 
\begin{align}
    p(x)=q\int_0^x\frac{\mathrm{d}s}{k(s)}.
\end{align}
We cannot obtain an analytic solution for $u$ for general $\kappa_1(x)$ and $\bar{\gamma}(x)$. %However, in the following sections we consider specific relevant cases of interest. 

%When $\bar{\gamma} = 0$ it is again straightforward to find an exact solution of this problem (see Appendix F). \newtext{However, for a general $\bar{\gamma}\newtext{(x)} \neq 0$ (and general $\kappa_1(x)$), \eqref{DarcyInhomogeneous} is a first-order nonlinear equation for ${\D{u}}/{\D{x}}$ that, to the best of our knowledge, does not possess an analytic solution. In order to make analytic progress, in the following sections we will separately consider two constitutive assumptions on the operating permeability from \eqref{eq:k img dimensionless}, namely
%\begin{equation}\label{eqn:const_1}
%k = \kappa_1(x) + \bar{\gamma} \frac{\D{u}}{\D{x}},
%\end{equation}
%where $\bar{\gamma}$ is independent of $x$, and \begin{equation}\label{eqn:const_2}
%k = \kappa_1(x) \left( 1 + \delta \frac{\D{u}}{\D{x}} \right),
%\end{equation}
%where $\delta$ is independent of $x$. For each of these assumptions we will study industrially relevant questions as outlined in the introduction to this section.
%\subsection{Constitutive assumption %\texorpdfstring{\eqref{eqn:const_1}}{eqn:const1}}
%As we are primarily interested in studying the effects of nonuniform starting permeability, the most natural simplifying assumption to make on the operating permeability is that $\tilde{k}_2$ is independent of $\tilde{x}$, which means that $\bar{\gamma}\newtext{(x)} \equiv \bar{\gamma}$ is constant too and we arrive at \eqref{eqn:const_1}.} 

%%%%%%%%%%%%%%%%%%%%%%%%%%%%%%%%%%%%%%%%%%%%%%%%%%%%%%%%%

\subsection{Uniform operating permeability}
\label{subsub:uniform_post}
We first seek a starting filter permeability function $\kappa_1(x)$ such that, when the filter is deformed under the action of fluid pressure gradients, the resulting permeability of the filter is uniform. This is of considerable industrial interest, since such a filter offers uniform removal properties throughout its depth and would negate the effects of compression during operation. Mathematically, this amounts to finding $\kappa_1(x)$ such that
\begin{equation}\label{ConstPermAfter}
 \kappa_1(x) + \bar{\gamma}(x) \frac{\D{u}(x)}{\D{x}} = k_{\mathrm{uni}},  
\end{equation}
using \eqref{eq:k img dimensionless}, where $k_{\mathrm{uni}}$ is a constant. Using \eqref{DarcyInhomogeneous}, \eqref{BC: u,p img} and \eqref{ConstPermAfter} it is straightforward to show that
\begin{eqnarray}
&q=k_{\mathrm{uni}}, \qquad u(x)=\displaystyle\frac{x^2}{2}-x, \qquad p(x)=x, \\[2mm]
&\kappa_1(x)+\bar{\gamma}(x)(x-1)=k_{\mathrm{uni}}, \\[2mm]
&k_{\mathrm{uni}}=1+\displaystyle\int_0^1 \bar{\gamma}(x)(x-1)\,\mathrm{d}x. 
\end{eqnarray}

To determine the starting permeability explicitly we must specify a relationship for $\bar{\gamma}(x)$. While this will depend on the particular filter material, there are two natural choices. Firstly, we could choose $\bar{\gamma}=\delta\kappa_1(x)$, where $\delta$ is a constant. In this case the permeability changes under compression in a way that is proportional to the local rest-state permeability.   
Secondly, we could choose $\bar{\gamma}=$\,constant. This corresponds to a medium whose permeability changes in a way that is directly proportional to its compression, regardless of its rest-state permeability. 

In Case (i) we obtain 
\begin{align}
\label{eq:case i uniform permeability}
 \kappa_1(x) &= \displaystyle \frac{-\delta}{(1+ \delta (x-1)) \ln{\left( {1-\delta} \right)} }, & k_{\mathrm{uni}} = \displaystyle \frac{-\delta}{\ln{\left( {1-\delta} \right)}},
\end{align}
while in Case (ii) we find that 
\begin{align}
\label{eq:case ii uniform permeability}
\kappa_1(x)&=1-\bar{\gamma}\left(x-\frac{1}{2}\right), &
k_{\mathrm{uni}}&=1-\frac{\bar{\gamma}}{2}.
\end{align}
%\begin{eqnarray}\label{eqn:linear_starting_perm_gives_uniform}
%q = k_{\mathrm{uni}} = 1 - \frac{\bar{\gamma}}{2},&   & \kappa_1(x) =  1 - \bar{\gamma}\left(x-\frac{1}{2}\right), \\ 
%u(x) =  \frac{x^2}{2} - x   &   & p(x) = x.
%\end{eqnarray}
Thus, in Case (ii), for a filter with a given value of $\bar{\gamma}$, the linear starting permeability with gradient $-\bar{\gamma}$ results in a spatially uniform permeability of $1- {\bar{\gamma}}/{2}$ under operation. We note that in this case the flux $q$ achieved for such a filter is equal to that achieved by a filter with an initially uniform permeability ($\kappa_1=1$) (see~\eqref{eq:Compression_NonDim_QDependentOnP0}).

%It follows that for a filter with a given value of $\delta$, a nonlinear starting permeability of the form \eqref{eq:case ii uniform permeability} \textcolor{red}{change this to a (b)} results in a spatially uniform permeability of $C(\delta)$ under operation.

%%%%%%%%%%%%%%%%%%%%%%%%%%%%%%%%%%%%%%%%%%%%%%%%%%%%%%%%%%%%%%%%%%%%%%%%%%%%%%%%%%%%%%%%%%%%%%%%%%%%%%%%%%%%%%

\subsection{Maximizing the flux}
\label{A general linear rest-state permeability}

We now turn our attention to the question of how we should choose the rest-state permeability to maximize the flux during operation for each of the two permeability relationships described in the previous section. \\[3mm]

\subsubsection{\textbf{Case (i) $\mathbf{k=\kappa_1(x)(1+\delta\mathrm{d}u/\mathrm{d}x)}$}}

Dividing both sides of \eqref{DarcyInhomogeneous} by $\kappa_1(x)$ %, we get
%\begin{equation}
%\frac{q}{\kappa_1(x)} = \frac{\DSS{u}}{\DS{x}} \left( 1 + \delta \frac{\D{u}}{\D{x}} \right),    
%\end{equation}
%which 
and integrating, using boundary conditions (\ref{BC: u,p img}b) and \eqref{eq:strain_at_fixed_end_rescaled} yields
%\begin{equation}
%q \int\limits_0^1 \frac{\D{x}}{\kappa_1(x)} = \left[ \frac{\D{u}}{\D{x}} + \frac{\delta}{2} \left( \frac{\D{u}}{\D{x}}\right)^2 \right]_0^1 =  1 - \frac{\delta}{2}
%\end{equation}
%and we conclude 
\begin{equation}
q = \displaystyle \frac{1- \frac{\delta}{2}}{ \int\limits_0^1 \frac{\D{x}}{\kappa_1(x)}}.    
\end{equation}
It follows from the Cauchy-Schwarz inequality that, for any continuous $\kappa_1(x)$ with a mean of $1$,
\begin{equation}
\displaystyle \frac{1}{\int\limits_0^1 \frac{\D{x}}{\kappa_1(x)}} \leq \int\limits_0^1 \kappa_1(x) \D{x} = 1
\end{equation}
and the equality holds if and only if $\kappa_1(x) \equiv 1$. This implies that for any $0< \delta <1$ we have
\begin{equation}
q \leq 1- \frac{\delta}{2}
\end{equation}
and the flux $q$ achieves its maximum value, $1-\delta/2$, for (and only for) an initially uniform rest-state permeability $\kappa_1(x)=1$.
Thus we find that we cannot choose an initial rest state that both leads to a uniform operating permeability (satisfied by choosing $\kappa_1$ according to \eqref{eq:case i uniform permeability}) and maximizes the flux (satisfied by choosing $\kappa_1=1$).\\

\subsubsection{\textbf{Case (ii) $\mathbf{k=\kappa_1(x)+\bar{\gamma}\mathrm{d}u/\mathrm{d}x}$}}
In this case it is not possible to make analytic progress for the most general form of $\kappa_1(x)$. However, since we have shown that, for a given operating regime, a linear rest-state permeability leads to a uniform permeability under operation we consider arbitrary linear rest-state permeability distributions of the form 
\begin{equation}
\label{eq:linear permeabilities}
\kappa_1(x) = 1 + \alpha \left(x-\frac{1}{2}\right),
\end{equation}
with the view of determining the permeability gradient $\alpha$ that maximizes the flux. Here we assume $0< \vert \alpha\vert < 2$, so that the rest-state permeability is everywhere positive (the results for $\alpha = 0$ were obtained in Section \ref{sec:filter_uniform}).   

We first write equation \eqref{DarcyInhomogeneous} in terms of the permeability $k$, 
\begin{equation}\label{equation_in_k}
k \left( \frac{\mathrm{d}k}{\mathrm{d}x} - \alpha\right) = q \bar{\gamma}.
\end{equation}
The boundary conditions \eqref{BC: u,p img} can be expressed in terms of $k$ as 
{\refstepcounter{equation}
\[
\label{opt_control_BCs_in_k}
 k(0) = 1- \frac{\alpha}{2} - \bar{\gamma}, \qquad k(1)  = 1 + \frac{\alpha}{2}.
\eqno{(\theequation\textrm{a,b})}
\]}

Under the assumption of linear starting permeability, if $\mathrm{d}k/\mathrm{d}x$ is positive/zero/negative at $x=0$, then $\mathrm{d}k/\mathrm{d}x$ is positive/zero/negative for all $x \in [0,1]$ (see Appendix C for details). It therefore follows that either the operating permeability $k$ is uniform in space or $\mathrm{d}k/\mathrm{d}x \neq 0$ for any $x \in [0,1]$. The former is only possible if $k(0)=k(1)$, which (using \eqref{opt_control_BCs_in_k}) is equivalent to $\alpha = - \bar{\gamma}$ and is thus consistent with our result from \eqref{eq:case ii uniform permeability}. If $\alpha \neq - \bar{\gamma}$, we then have $\mathrm{d}k/\mathrm{d}x \neq 0$ everywhere, and \eqref{equation_in_k} can be inverted as
\begin{equation}
\label{eq:img dx/dk}
\frac{\D{x}}{\D{k}} = \frac{k}{q \bar{\gamma} +\alpha k}.
\end{equation}

Integration of \eqref{eq:img dx/dk} and  application of the boundary conditions (\ref{opt_control_BCs_in_k}) gives the implicit relation for $k(x)$,
\begin{align}\label{implicit_for_q}
    x&=\frac{k}{\alpha}-\frac{q\bar{\gamma}\ln(q\bar{\gamma}+k\alpha)}{\alpha^2}+\frac{1}{2}+\frac{\bar{\gamma}-2}{2\alpha} \\ &+\frac{q\bar{\gamma}}{2\alpha^2}\ln(q\bar{\gamma}+\alpha-\alpha^2/2  -\alpha\bar{\gamma})+\frac{q\bar{\gamma}}{2\alpha^2}\ln(q\bar{\gamma}+\alpha+\alpha^2/2). \nonumber
\end{align}
The strain is then given from \eqref{eq:k img}: 
\begin{align}
\label{eq: u' sol img}
\frac{\mathrm{d}u}{\mathrm{d}x}=\frac{k-\kappa_1(x(k))}{\bar{\gamma}},
\end{align}
 and the deformation is given by integrating this and using \eqref{eq:img dx/dk} and \eqref{eq: u' sol img},
\begin{align}
    u=\int_{1-\alpha/2-\bar{\gamma}}^k \frac{s-1-\alpha(x(s)-1/2)}{\bar{\gamma}(q\bar{\gamma}/s+\alpha)}\,\mathrm{d}s.
\end{align}

Using the fact that we require $q>0$ for physical solutions and boundary conditions (\ref{opt_control_BCs_in_k}), we obtain from \eqref{implicit_for_q} the implicit relation between $q$, $\alpha$ and $\bar{\gamma}$,
%\begin{equation}\label{implicitly_q}
%\alpha = q \ln{\left(1 + \frac{\alpha^2 + \alpha \bar{\gamma}}{q \bar{\gamma} + \alpha - \alpha^2/2 - \alpha \bar{\gamma}} \right)},
%\end{equation}
%Physically relevant solutions require $q>0$
\begin{equation}
\label{eq:img gamma sol}
q(\alpha, \bar{\gamma}) \ln \left(1 + \frac{\alpha ^2+ \alpha  \bar{\gamma} }{-\frac{\alpha ^2}{2}- \alpha  \bar{\gamma} +\alpha + \bar{\gamma}  q(\alpha, \bar{\gamma})} \right)-\alpha = 0,
%\bar{\gamma}= -\frac{\alpha  \left(\alpha +(\alpha -2) e^{\frac{\alpha }{q}}+2\right)}{2 \left(e^{\frac{\alpha }{q}} (\alpha -q)+q\right)}.
\end{equation}
which holds for $\alpha \neq 0$ and $ \alpha \neq - \bar{\gamma}$ (with these two subcases addressed in Sections~\ref{sec:filter_uniform} and \ref{subsub:uniform_post} respectively).
%q \log \lefgammat(.\frac{\alpha ^2+\ala \neq - \bar{\}pha  \gamma }{-\frac{\alpha ^2}{2}-\alpha  \gamma +\alpha +\gamma  q}+1\right)-\alpha
Assuming that we can find a unique solution $q(\alpha,\bar{\gamma})$ for given $\alpha$ and $\bar{\gamma}$ in certain ranges (for a detailed description of such parameter ranges and proof of the existence and uniqueness of solutions see Appendix D), we can differentiate this equation with respect to $\alpha$, setting $\partial q/ \partial \alpha = 0$ and solve for $q$. One concludes that the maximum flux $q_{\mathrm{max}}$ is achieved for a value $\alpha=\alpha_{\mathrm{max}}(\bar{\gamma})$ that satisfies 
\begin{align}
&q_{\mathrm{max}}(\bar{\gamma}) = q(\alpha_{\mathrm{max}}(\bar{\gamma}),\bar{\gamma}) = \frac{1}{16\bar{\gamma}} \bigg[{2 \alpha_{\mathrm{max}}(\bar{\gamma})  (\bar{\gamma} -2)}-4 \bar{\gamma}  (\bar{\gamma} -2) \nonumber \\ &
+ \Big( 4 (\bar{\gamma} -2)^2 (\alpha_{\mathrm{max}}(\bar{\gamma}) -2 \bar{\gamma} )^2 \label{eq:img maximum flux} \\ 
& -32 \alpha_{\mathrm{max}}(\bar{\gamma})  (\alpha_{\mathrm{max}}(\bar{\gamma}) +2) \bar{\gamma}  (\alpha_{\mathrm{max}}(\bar{\gamma}) +2 \bar{\gamma} -2) \Big)^{1/2} \bigg]. \nonumber
\end{align}
Substituting $q_{\mathrm{max}}$ from \eqref{eq:img maximum flux} into \eqref{eq:img gamma sol} provides an implicit equation for $\alpha_{\mathrm{max}}$ for a given $\bar{\gamma}$; the corresponding flux is then given by \eqref{eq:img maximum flux} for that value of $\bar{\gamma}$.

\begin{figure}
\centering
\subfloat[]{\begin{overpic}[width=0.23\textwidth,tics=10]{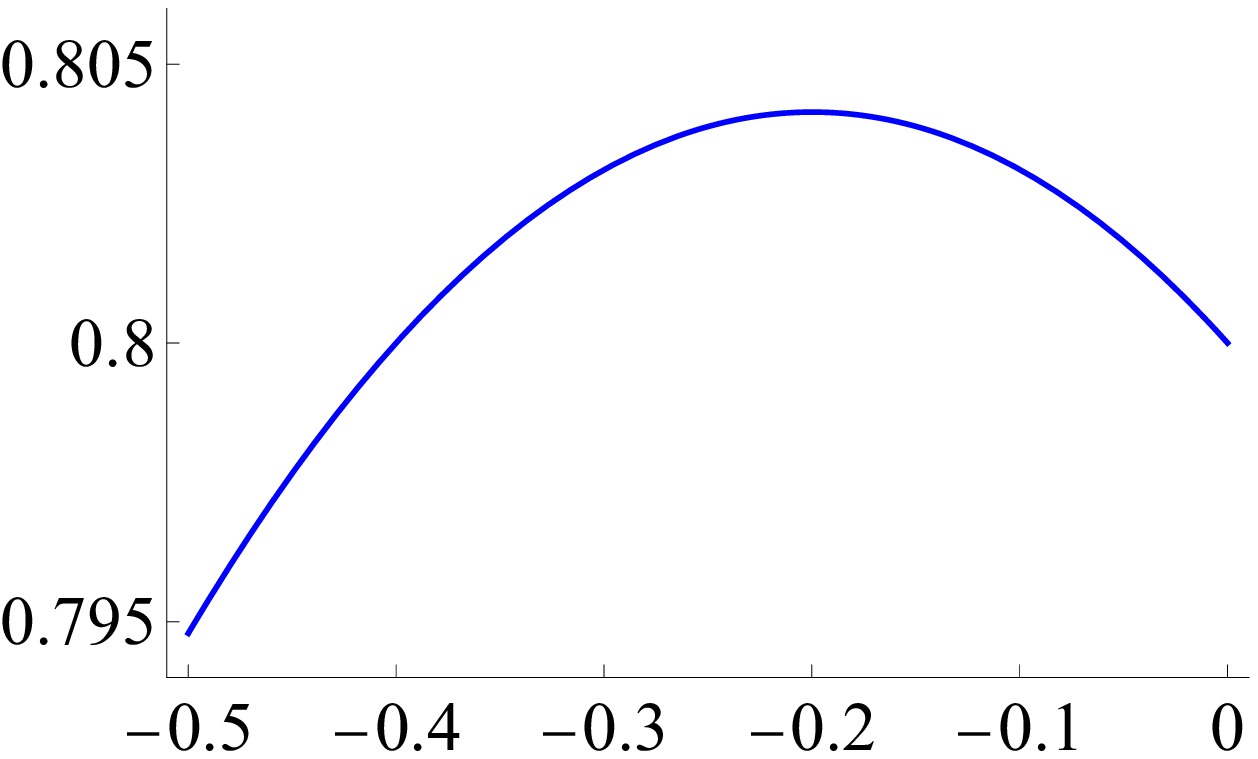}
{\normalsize \put(103,-5){$\alpha$}
\put(12,75){$q$}
\put(60,65){\color{blue} $\alpha = \alpha_{\mathrm{max}}$}
\put(19,52){\color{red} $\alpha = - \bar{\gamma}$}
\put(36,38){\color{red} \circle*{4.0}}
\put(111,38){\color{black} \circle*{4.0}}
\put(73,59){\color{blue} \circle*{4.0}}
\put(15,38){\color{black} \linethickness{0.03cm}\line(1,0){100}}
\put(45,25){\color{black} $q = 1- \bar{\gamma}/2$}}
\end{overpic}}
\subfloat[]{\begin{overpic}[width=0.23\textwidth,tics=10]{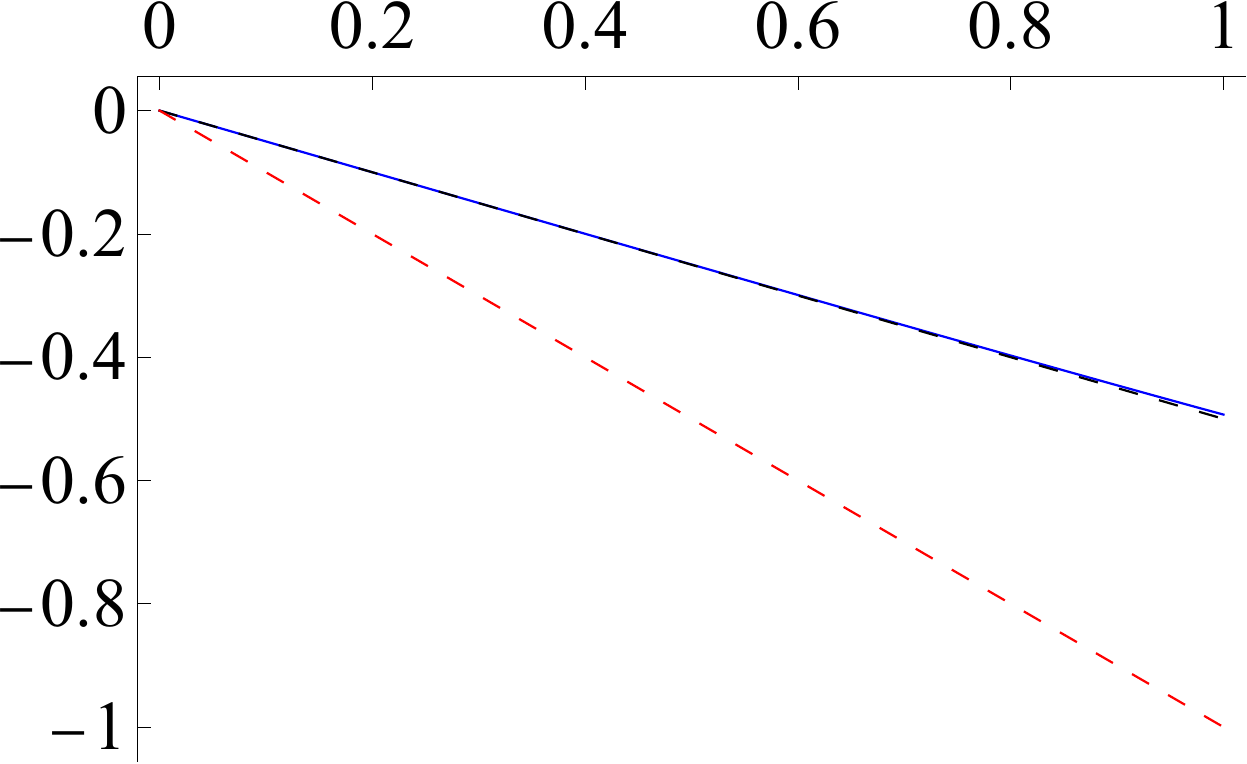}
\put(10,-10){$\alpha$}
\put(105,52){$\bar{\gamma}$}
\put(80,21){$\alpha = - \bar{\gamma}/2$}
\put(75,45){\color{blue}$\alpha_{\mathrm{max}}$ \color{black}}
%\put(175,90){\color{green} $\alpha = \alpha_{\mathrm{max}}^{(2)}(\bar{\gamma})$}
\put(25,25){\color{red}$\alpha = - \bar{\gamma}$}
\end{overpic}}
 \caption{(a) The dependence of flux $q$ on $\alpha$ for $\bar{\gamma} = 0.4$. (b)~The value of $\alpha$, $\alpha_{\mathrm{max}}$, that maximizes the flux for a given $\bar{\gamma}$. Note $\alpha_\textrm{max}$ is very close to $-\bar{\gamma}/2$ (plotted in dashed black for comparison). The value of $\alpha$ required to achieve uniform permeability under operation is shown for comparison by the red dashed line.}
\label{fig:implicit_alpha_max}
\end{figure}

We plot the variation of the flux versus $\alpha$ (found numerically) for $\bar{\gamma}=0.4$, showing: the maximum flux when  $\alpha=\alpha_{\mathrm{max}}$; the flux when the permeability is uniform everywhere, $\alpha=- \bar{\gamma}$, found in Section~\ref{subsub:uniform_post}; and the flux when we start with a spatially uniform permeability, $\alpha=0$ in Figure~\ref{fig:implicit_alpha_max}(a).

We observe that $\alpha_{\mathrm{max}}$ is remarkably close to $-\bar{\gamma}/2$ (Figure \ref{fig:implicit_alpha_max}b). Indeed, if we consider the small-$\bar{\gamma}$ limit we find $\alpha_{\mathrm{max}} \approx -\bar{\gamma}/2+O(\bar{\gamma}^2)$. However, the solution does diverge with increasing $\bar{\gamma}$, albeit slowly. This choice of $\alpha$ corresponds to one half of the value that yields uniform permeability post-flow ($\alpha=-\bar{\gamma}$, see Equation \eqref{eq:case ii uniform permeability}). This indicates that, as found for a material with a permeability relationship $\kappa(x)=\kappa_1(x)(1+\delta\mathrm{d}u/\mathrm{d}x)$ (Case (i)), for a material whose permeability varies according to $\kappa(x)=\kappa_1(x)+\bar{\gamma}\mathrm{d}u/\mathrm{d}x$ (Case (ii)) we cannot fabricate a filter with an initial linear permeability distribution that both possesses a uniform permeability distribution under operation (satisfied by choosing $\kappa_1=1+\alpha(x-1/2)$ with $\alpha=-\bar{\gamma}$) and maximizes the flux (satisfied by choosing $\kappa_1=1+\alpha(x-1/2)$ with $\alpha=\alpha_{\mathrm{max}}$ given by  \eqref{eq:img maximum flux} into \eqref{eq:img gamma sol} and shown in Figure \ref{fig:implicit_alpha_max}b).

%%%%%%%%%%%%%%%%%%%%%%%%%%%%%%%%%%%%%%%%%%%%%%%

\iffalse
\subsection{Constitutive assumption \texorpdfstring{\eqref{eqn:const_2}}{eqn:const2}}

Provided, one wishes to keep both $\tilde{k}_1$ and $\tilde{k}_2$ dependent on $\tilde{x}$, the simplest assumption to make (and one using which we can still solve the problem analytically) is that $\tilde{k}_2 = \delta \tilde{k}_1$, where $\delta$ is independent of $\tilde{x}$. This way, one arrives at \eqref{eqn:const_2}. Note that the problem is well-posed provided $\kappa_1(x)>0$ everywhere and $0<\delta<1$.

\subsubsection{Uniform operating permeability}

Similar to Section \ref{subsub:uniform_post}, one imposes
\begin{equation}\label{ConstPermAfter_2}
 \kappa_1(x) \left( 1 + \delta \frac{\D{u}}{\D{x}} \right) = k_{\mathrm{uni}}, 
\end{equation}
and upon simple analysis concludes
\begin{equation}\label{eqn:linear_starting_perm_gives_uniform_2}
q = k_{\mathrm{uni}} = \displaystyle \frac{\delta}{\ln{\left( \frac{1}{1-\delta} \right)}} \quad \text{and} \quad \kappa_1(x) = \displaystyle \frac{\delta}{(1+ \delta (x-1)) \ln{\left( \frac{1}{1-\delta} \right)} }.
\end{equation}
It follows that for a filter with a given value of $\delta$, a nonlinear starting permeability of a form $C(\delta)/(1+\delta(x-1))$ where $C(\delta) = - \delta/(\ln{(1-\delta)})$, results in a spatially uniform permeability of $C(\delta)$ under operation.
\fi 

%%%%%%%%%%%%%%%%%%%%%%%%%%%%%%%%%%%%%%%%%%%%%%%%%%%%%%%%%%%%%%%%%%%%%%
\section{Conclusions}

\label{Section:Conclusions}

In this paper we have outlined a mathematical model for the behaviour of a filter during operation that takes into account its compressibility. Specifically, when a pressure difference is applied to a porous medium, the material will compress in a non-uniform manner, which in turn will change the permeability. Equations for linear elasticity, Darcy's law and mass conservation were employed to describe the system. 

We began by considering a porous medium composed of a uniform material, with a corresponding spatially uniform permeability in its rest state, that is, when no pressure is applied. We considered the case where a portion of this material was attached to a highly porous fixed grid. When fluid was driven  through the medium we found that it compressed in a non-uniform manner, with the level of compression increasing as we moved from the free end to the end held in place by the grid. We proposed a simple linear law relating the level of compression to the permeability of the medium. We found that, as the applied pressure difference was increased, the flux through the medium increased along with the level of compression. Further, we obtained analytic expressions that describe the entire system, with the behaviour characterized by a single parameter that provides a measure of the compressibility of the material for a given applied pressure. Eventually, as the applied pressure increases or the material elasticity increases, a critical value of the parameter is attained at which the permeability reaches zero and fluid can no longer pass through the medium. The applied pressure difference just below this critical point is found to generate the highest flux. Expressing the system in terms of this single parameter allowed us to quantify how the highest flux depends on the operating regime and material properties.

In a typical membrane filtration we care not only about the flux but also the amount of power required to generate that flux. We defined a metric that balanced the desire to maximize flux while minimizing the power required and, for a given weighting, analytically determined the value of our single system parameter that optimized the operation. 

We then turned our attention to the idea of using a filter with a spatially varying permeability, motivated by the fact that such tailored materials can improve performance. We considered two types of material: one whose permeability changed upon compression either independently or proportionally to the local rest-state permeability.  

We first observed that by choosing a filter with a linear permeability distribution in its rest state we could obtain a filter with uniform permeability under operation. Such filters behave uniformly with depth and so can lead to more desirable contaminant trapping. We analytically determined the permeability gradient required to achieve this for a given applied pressure.

We then expanded our search to the general class of linear-permeability rest-state filters with the view of improving the flux performance. We analytically determined the compressive behaviour of such a filter under an applied pressure difference and found the permeability gradient required to maximize the flux. While one might expect that the optimal permeability gradient would be one that also leads to a uniform permeability under operation, we found that we could not choose one rest-state permeability that satisfies both properties. 

The ideas that we have outlined here form a simple basis for viewing the behaviour of a filter under operation and highlights the importance of considering the compressibility, despite its lack of attention to date. The ideas form a fundamental basis for studying the removal properties of a filter under operation.

%%%%%%%%%%%%%%%%%%%%%%%%%%%%%%%%%%%%%%%%%%%%%%%%%%%%%%%%%%%%%%%%%%%%%%
\begin{acknowledgment}

JK is grateful to the Royal Society for financial support. IMG gratefully acknowledges support from the Royal Society through a University Research Fellowship. AUK was funded by EPSRC. 

\end{acknowledgment}

\appendix       %%% starting appendix

\section*{Appendix A: Non-applicability of the model for $\mathbf{\gamma}\mathbf{>1}$}

Here we show that for $\gamma > 1$, the dimensionless model consisting of the Navier equation \eqref{eq:NonDimNavierEquation}, Darcy's law \eqref{eqn:Darcy_rescaled} with the permeability $k$ modified to allow for shutdown:
\begin{equation}\label{perm_modified}
k = \max \left\{ 1+ \gamma \frac{\D{u}}{\D{x}}, 0 \right\}    ,
\end{equation}
equipped with boundary conditions \eqref{eq:Compression_Rescaled_BC} possesses no solutions with continuous strain distribution $\D{u}/\D{x}$. Assume for contradiction that such a solution exists. Using the values of strain at the fixed and the free ends (see Equations (\ref{eq:Compression_Rescaled_BC}b) and \eqref{eq:strain_at_fixed_end_rescaled}) and the assumption of its continuity we conclude that there must exist an $x^* > 0$ such that $k=0$ for all $x<x^*$ (we pick $x^*$ to be the supremum of all $x$ satisfying this property), and also $x^{**}<1$ such that $k>0$ for all $x>x^{**}$ (we pick $x^{**}$ to be the infimum of all $x$ satisfying this property and from continuity, we get that $k (x=x^{**}) = 0$ and thus $\D{u}/\D{x} (x=x^{**}) \leq -1/\gamma$). As $k=0$ for all $x \in (0,x^{*})$, we conclude that the flux through the medium must be zero and Darcy's law then tells us that $p$ must be constant for $x \in [x^{**},1]$. The Navier equation then implies that $\D{u}/\D{x}$ is constant in $[x^{**},1]$, which gives $\D{u}/\D{x} (x=x^{**}) = 0$. This is in contradiction with the previously obtained result $\D{u}/\D{x} (x=x^{**}) \leq - 1/\gamma$.
\\
Note that the expression for strain as given by \eqref{eq:ClosedFormOfPTuPTx_2} thus does not hold for $\gamma>1$, the reason being that in the solution process we used the Navier equation to eliminate the pressure variable (the $\D{p}/\D{x}$ term) from Darcy's law, but we did not account for the pressure boundary conditions. For the same reasons, the expression \eqref{eq:ClosedFormOfPTuPTx_2} does not satisfy \eqref{eq:strain_at_fixed_end_rescaled}.
\\
In the notation from \cite{Parker1987}, for $p_1 > \hat{R}$, expression (3.12) for deformation $R$ evaluated at $x=0$ does not satisfy the strain boundary condition at the fixed end $x=0$ (3.9). Consequently, the positive flux $w$ for $p_1 > \hat{R}$ as given by expression (3.13) (and as plotted in Fig. 3) is not a correct outcome of the model as it stands. We also note that the decline of the flux beyond the flux-maximizing value of pressure gradient as predicted in \cite{Parker1987} (Figure 3) has not been observed in the experiments performed in the same work (Figure 9).

\section*{Appendix B: Finding the maximizers in \textbf{(}\textbf{\ref{F_alpha_gamma_opt}}\textbf{)}}
Setting $\mathcal{F}_{\xi}'(\gamma^*)=0$, where a prime denotes differentiation, yields the stationary points
\begin{equation}\label{stationary_gamma}
\gamma^{*}(\xi) = \frac{2 (2 \xi -1)}{3 \xi -2}.    \end{equation}
To determine if these are maxima and whether they belong to the interval of interest ($\gamma \in [0,1]$), we split the analysis into three cases.
\\
\indent If $\xi \in [0,1/2]$, then $\mathcal{F}'_{\xi}(\gamma) \geq 0$ for any $\gamma < \gamma^*$ and $\mathcal{F}'_{\xi}(\gamma) \leq 0$ for any $\gamma > \gamma^*$, implying that $\gamma^*$ as defined in Equation \eqref{stationary_gamma} is the optimum value $\gamma_{\mathrm{opt}}$.
\\
\indent If $\xi \in [1/2, 2/3]$, then $\gamma (3 \xi -2) + 2 (1- 2 \xi) \leq 0$, which implies that $\mathcal{F}'_{\xi}(\gamma) \leq 0$ for any $\gamma$, $\mathcal{F}_{\xi}(\gamma)$ is a non-increasing function of $\gamma$ and thus $\gamma_{\mathrm{opt}}=0$.
\\
\indent If $\xi \in [2/3,1]$, then (recalling that $\gamma \leq 1$) we have $\gamma (3 \xi -2) + 2 (1- 2 \xi) \leq 3 \xi -2 + 2- 4 \xi  < 0$. As before, $\gamma_{\mathrm{opt}} = 0$ for such $\xi$.
\\
Thus, we have proven that for a given $\xi$, the value of $\gamma$ that maximizes $\mathcal{F}_{\xi}(\gamma)$ is that from \eqref{eqn:gamma_opt}.

\section*{Appendix C: Operating permeability under linear starting permeability must be either strictly monotonic or constant}

We differentiate \eqref{equation_in_k} with respect to $x$ and get
\begin{equation}
\left(k k' - \alpha k \right)' = k' (k' - \alpha) + k k'' = 0,
\end{equation}
which can be rewritten as a system of two first-order equations (assuming $k > 0$):
\begin{eqnarray}\label{dyn_system_2}
k' =& l, \\
\nonumber
l' =& \displaystyle\frac{l (\alpha - l)}{k}.
\end{eqnarray}
Clearly, for any $k>0$ $(k,0)$ is an equilibrium point of this dynamical system. Therefore, if $k'(0)=0$, then $k'(x)=0$ for all $x \in [0,1]$. Furthermore, if we assume (for the sake of contradiction) that there exists $x_* \in (0,1]$ such that $l(0) l(x_*) < 0$ (i.e.~$l$ has opposite signs at $0$ and $x_*$), then assuming continuity of $k' = l = \alpha + \bar{\gamma} u''$ (i.e.~assuming continuity of $u''$), there would have to exist $0<x_{**}<x_{*}$, for which $l(x_{**})=0$. As $l=0$ are equilibrium points of \eqref{dyn_system_2}, it would follow that $l(x)=0$ for all $x \in [x_{**},1]$ and therefore $l(x_*)=0$, which contradicts our assumption $l(0) l(x_*) < 0$.
\\
It remains to show that if $k'(0) \neq 0$, then $k'(x) \neq 0$ for all $x \in [0,1]$, from which it follows that $k'(x)$ is either positive or negative for all $x \in [0,1]$, and thus the operating permeability $k(x)$ must be strictly monotone.
\\
If $2> \alpha > 0$, then for $0< k'(0) = l(0) < \alpha$, we have $l'(x) \geq 0$ (implying $k'(x) = l(x) > 0$) for all $x \in [0,1]$, and for $0 > k'(0)=l(0)$, we have $l'(x) < 0$ (implying $k'(x) = l(x) < 0 $ for all $x \in [0,1]$).
\\
It follows from our observations that $k$ cannot decrease in one part of the domain and increase in another and thus the global %maxima and
minima of $k$ must always be attained at the boundaries, which means that we get, using the boundary conditions \eqref{opt_control_BCs_in_k}, %$$k_{\mathrm{max}}(\alpha, \bar{\gamma}): = \max\limits_{x %\in [0,1]} k(x) = \max \left\{ 1-\alpha/2 - \bar{\gamma} , 1 %+ \alpha/2 \right\}$$
%and
$$k_{\mathrm{min}}(\alpha, \bar{\gamma}) := \min\limits_{x \in [0,1]} k(x) = \min \left\{ 1-\alpha/2 - \bar{\gamma} , 1 + \alpha/2 \right\}.$$
Consequently, for any fixed $-2 < \alpha \leq 0$ and $0\leq \bar{\gamma}<1 - \alpha /2$ (for which we show the well-posedness of our problem in Appendix D), we get $0<k_{\mathrm{min}}(\alpha, \bar{\gamma}) \leq k(x)$ for any $x \in [0,1]$.
\\
If $y_0 := l(0)>0$, then we have $l'(x) \geq l(x)(\alpha - l(x))/k_\mathrm{min}$. The solution to $y' = y(\alpha - y)/k_\mathrm{min}$ equipped with $y(0)=y_0$ reads 
$$y(x) = \frac{\alpha y_0 e^{\frac{\alpha}{k_\mathrm{min}} x}}{\alpha - y_0 + y_0 e^{\frac{\alpha}{k_\mathrm{min}}x}}$$
for $\alpha \neq 0$ and
$$y(x) = \frac{k_\mathrm{min} y_0}{k_\mathrm{min} + x y_0}$$
for $\alpha=0$. In either case, it follows that $y(x)$ is positive for all finite $x$ (and thus for all $x \in [0,1]$). Given that we have $l(0)=y(0)$ and the right-hand side of the equation for $l$ is bounded from below by the right-hand side of the equation for $y$, $y(x)$ acts as a positive subsolution, and therefore we conclude that $l(x) > 0 $ for all $x \in [0,1]$.
\\
The same arguments apply for $\alpha< l(0)<0$ (here we need to bound $l'$ from above and $y(x)$ acts as a negative supersolution) and for $l(0) \leq \alpha \leq 0$, we observe that $l'(x) \leq 0$ for all $x \in [0,1]$. In both cases we conclude (for $l(0)<0$) that $k'(x) = l(x)<0$ for all $x \in [0,1]$.
\color{black}

\section*{Appendix D: Well-posedness of the problem with linear starting permeability}

\begin{figure}
\centering
\subfloat[Region of well-posedness]{\begin{overpic}[width=0.24\textwidth,tics=10]{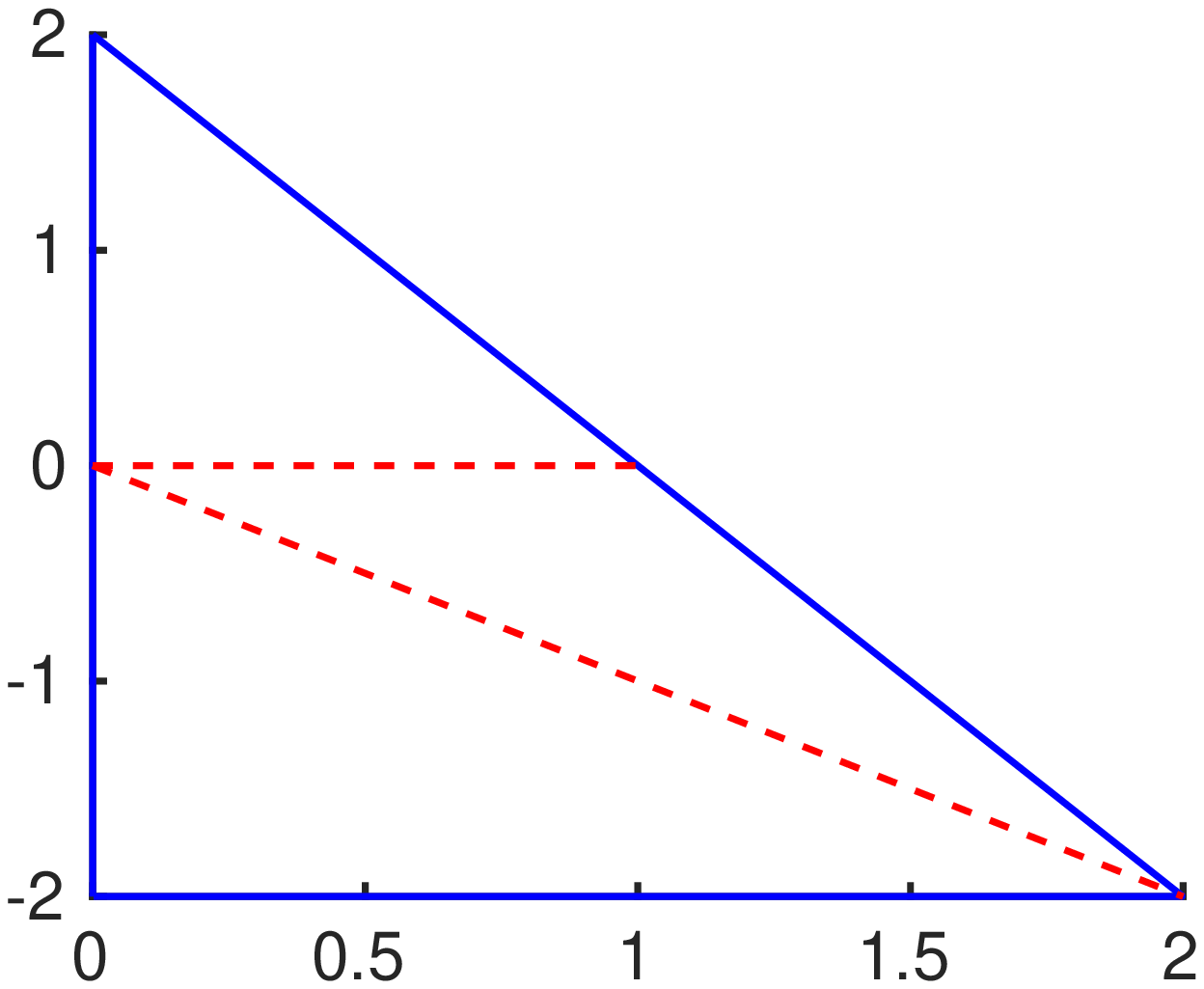}
%\linethickness{1pt}
\normalsize{
\put(95,0){$\bar{\gamma}$}
\put(12,88){$\alpha$}
\put(62,52){\color{blue}$\alpha = 2 - 2 \bar{\gamma}$}
\put(24,22){\color{red}$\alpha = - \bar{\gamma}$}
}
%\put(243,10){$\tilde{L}_d$}
%put(330,70){\vector(-1,0){70}}
%\put(278,122){\textbf{Flow}}
\end{overpic}}
\subfloat[Flux as function of $\bar{\gamma}$]{\begin{overpic}[width=0.24\textwidth,tics=10]{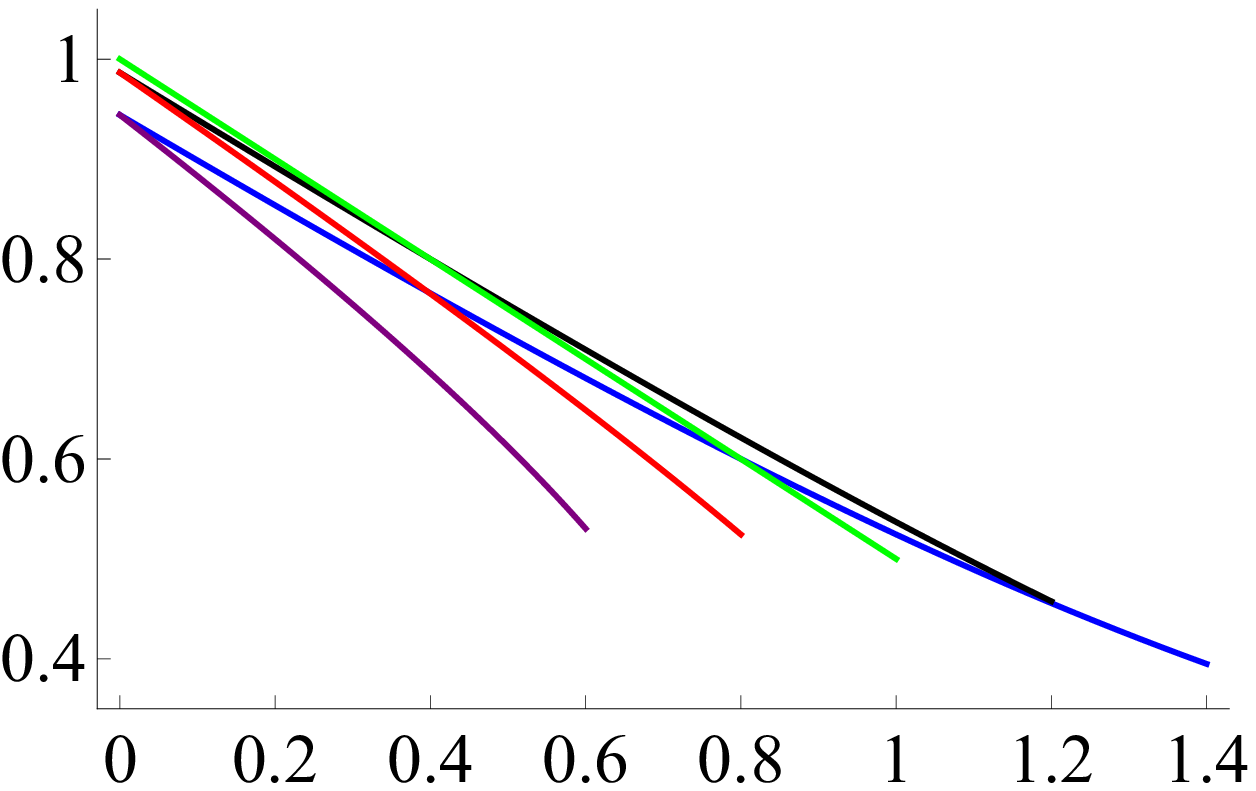}
\put(105,-8){$\bar{\gamma}$}
\put(7,80){$q$}
%\put(30,85){\color{blue} $\alpha = -0.1$}
%\put(15,50){\color{purple} $\alpha = -0.9$}
%\put(250,15){$\alpha$}
%\put(35,187){$q$}
%\put(145,187){\color{blue} $\alpha = \alpha_{\mathrm{max}}$}
%\put(65,187){\color{red} $\alpha = - \bar{\gamma}$}
%\put(78,92){\color{red} \circle*{6.0}}
%\put(232,92){\color{black} \circle*{6.0}}
%\put(156,124){\color{blue} \circle*{6.0}}
%\put(38,92){\color{yellow} \linethickness{0.03cm}\line(1,0){192}}
%\put(80,74){\color{yellow} $q = 1- \bar{\gamma}/2$}

\end{overpic}}

\caption{(a) The blue triangle represents the region in the $(\bar{\gamma}, \alpha)$ plane in which we expect the solution to exist (and be unique). The dashed red lines represent $\alpha = 0$ and $\alpha = - \bar{\gamma}$. (b) The dependence of flux $q$ on $\bar{\gamma}$ for $\alpha = -0.8$ (blue), $-0.4$  (black), 0 (green), 0.4 (red), and 0.8 (purple). Note that different values of $\alpha$ allow different ranges of $\gamma$ for which the solution exists (i.e. for which the operating permeability is positive everywhere; for details see Appendix D) and that for a fixed $\alpha$, $q$ is maximized when $\bar{\gamma}=0$.}
\label{fig:region_alpha_gamma}
\end{figure}

Recall that for $\alpha = 0$ we found the solution to our model for $0 \leq \bar{\gamma} < 1$ (avoiding $k=0$ at $x=0$). In the $\alpha > 0$ case, it suffices to ensure that $k(x=0)>0$, which gives $\alpha < 2 - 2 \bar{\gamma}$. For $\alpha <0$, Appendix C implies that if $\alpha > - \bar{\gamma}$, then the operating permeability is strictly increasing in $x$ and we thus expect the solution to exist whenever $k(x=0)$ is positive (i.e. $\alpha < 2 - 2 \bar{\gamma}$). Analogously, if $\alpha < -\bar{\gamma}$, we get that the operating permeability is strictly decreasing in $x$ and we expect the solution to exist whenever $k(x=1)$ is positive, which is equivalent to $\alpha > -2$. In summary, we expect the solution to exist in a region $\left\{ (\bar{\gamma}, \alpha) \in \mathbb{R}^2, 0 \leq \bar{\gamma} < 2, -2 < \alpha < 2 - 2 \bar{\gamma} \right\}$ (i.e.~inside the blue triangle drawn in Figure \ref{fig:region_alpha_gamma}a).

It follows from \eqref{target} that the flux $q$ must satisfy
\begin{eqnarray}
q^{-1} = && \int\limits_0^1 \frac{\D{x}}{1 + \alpha(x-1/2) + \bar{\gamma} \frac{\D{u}}{\D{x}}} \geq \nonumber \\ 
&& \int\limits_0^1 \frac{\D{x}}{1 + \alpha(x-1/2)} = \frac{1}{\alpha} \ln{\left( \frac{2+\alpha}{2-\alpha} \right)},
\end{eqnarray}
and so we obtain the bounds
\begin{equation}\label{bound_for_q}
0 < q \leq \frac{\alpha}{\ln{\left( \frac{2+\alpha}{2-\alpha} \right)}}=: q_{\bar{\gamma}=0}(\alpha).
\end{equation}
In Figure~\ref{fig:region_alpha_gamma}(b) we show the variation of $q$ as a function of $\bar{\gamma}$ for different values of $\alpha$, which shows that the maximum flux is achieved when $\bar{\gamma}=0$.

Recalling that equation~\eqref{eq:img gamma sol} was derived under the assumptions $\alpha \neq \bar{\gamma}$ and $\alpha \neq 0$, note that the substitution of $\alpha = -\bar{\gamma}$ (for any $\bar{\gamma} \in (0,2)$) into~\eqref{eq:img gamma sol} yields a contradictory result $\alpha = 0$ (unless the denominator inside the logarithm equals $0$, in which case we are dealing with a limit in an indeterminate form). Therefore, we rewrite equation~\eqref{eq:img gamma sol} in the form 
\begin{align}
\label{eq:mathcal L img}
\mathcal{L}(\alpha, \bar{\gamma}, q) := \left(q \bar{\gamma} + \alpha - \alpha^2 /2 - \alpha \bar{\gamma} \right) \left( e^{\frac{\alpha}{q}} -1  \right) - \alpha^2 - \alpha \bar{\gamma}=0
\end{align}
and observe that when $\alpha=-\bar{\gamma}$, \eqref{eq:mathcal L img} may be rearranged to give 
\begin{equation}
q = (2- \bar{\gamma})/2,   
\end{equation}
which implies that the solution found in Section \ref{subsub:uniform_post} also solves (and is the only solution of) \eqref{eq:mathcal L img} for $\alpha = -\bar{\gamma}$. However, $\mathcal{L}(0, \bar{\gamma}, q) = 0$ is still satisfied for any $q>0$ (and therefore does not define the flux uniquely) for a given $\bar{\gamma}$. Therefore, we study the existence and uniqueness of solutions $q \in (0, q_{\bar{\gamma}=0}(\alpha))$ to \eqref{eq:mathcal L img} in the parameter region $(\alpha,\bar{\gamma}) \in (-2,0) \times (0,(2- \alpha)/2)$ first, and then, separately in the region $(\alpha,\bar{\gamma}) \in (0,2) \times (0, (2- \alpha)/2)$. Note that the (only) solution in the $\alpha = 0$ case was explicitly found in Section \ref{Section:Solution}.

\subsection*{Existence}

Note first that for any $\alpha$ and $\bar{\gamma}$ of our interest, we have
\begin{eqnarray}\label{limit_L_upper_q} \displaystyle
& \lim\limits_{q \to q_{\bar{\gamma}=0}(\alpha)} \mathcal{L}(\alpha, \bar{\gamma},q) =  \frac{\alpha \bar{\gamma}}{\ln{(\Upsilon)}} \mathcal{R}(\Upsilon) ,
\end{eqnarray}
where we have introduced $\Upsilon:= (2+ \alpha)/(2- \alpha)$ and $\mathcal{R}(\Upsilon):= \Upsilon-1-\Upsilon \ln{(\Upsilon)}$. 

%\subsubsection*{{$\alpha \in (-2,0)$}{aIn-20}} 

First, fixing $\alpha \in (-2,0)$, we observe that $\Upsilon \in (0,1)$, $\mathcal{R}'(\Upsilon) = - \ln{(\Upsilon)} >0$ and $\lim\limits_{\Upsilon \to 1} \mathcal{R}(\Upsilon) = 0$. This means that $\mathcal{R}$ is negative in the region of our interest, and thus the same is true for $ \lim\limits_{q \to q_{\bar{\gamma}=0}(\alpha)} \mathcal{L}(\alpha, \bar{\gamma},q)$. Moreover
\begin{equation}
\lim\limits_{q \to 0} \mathcal{L}(\alpha, \bar{\gamma},q) = - \alpha \left( 1+ \frac{\alpha}{2} \right) >0.
\end{equation}
By the continuity of $\mathcal{L}$ as a function of $q$ (for $0<q<q_{\bar{\gamma}=0}(\alpha)$), we conclude that there must exist at least one $q \in (0, q_{\bar{\gamma}=0}(\alpha))$ for which $\mathcal{L}(\alpha, \bar{\gamma}, q) = 0$. 

%\subsubsection*{\texorpdfstring{$\alpha \in (0,2)$}{aIn02}}

When $\alpha \in (0,2)$, we have $\Upsilon \in (1, \infty)$ and $\mathcal{R}'(\Upsilon) = - \ln{(\Upsilon)} <0$, which together with $\mathcal{R}(1) = 0$ gives $\mathcal{R}(\Upsilon)<0$ and thus the right-hand side of \eqref{limit_L_upper_q} is again negative. Furthermore, one can also conclude that $\lim\limits_{q \to 0} \mathcal{L}(\alpha, \bar{\gamma},q) = \infty$ and the continuity of $\mathcal{L}$ again gives the existence of the desired $q$.

\subsection*{Uniqueness} 

We have seen that, in both regions of interest, $\lim\limits_{q \to 0} \mathcal{L}(\alpha, \bar{\gamma},q) >0$ and $\lim\limits_{q \to q_{\bar{\gamma}=0}(\alpha)} \mathcal{L}(\alpha, \bar{\gamma},q) < 0$. Thus, if we show that $\frac{\partial \mathcal{L}}{\partial q}(\alpha, \bar{\gamma},q) < 0$ for some $(\alpha, \bar{\gamma})$ and for any $q \in (0,q_{\bar{\gamma}=0}(\alpha))$, then we can conclude that for such $(\alpha, \bar{\gamma})$, the solution $q(\alpha,\bar{\gamma})$ of \eqref{eq:mathcal L img} (the existence of which was shown in the previous section) is unique (in $0< q < q_{\bar{\gamma}=0}(\alpha)$). Using $- \bar{\gamma} > (\alpha -2)/2$, we get
\begin{eqnarray} \nonumber
\displaystyle
\frac{\partial \mathcal{L}}{\partial q}(\alpha, \bar{\gamma},q) = &- \bar{\gamma} + \displaystyle\frac{2 \bar{\gamma} (q^2 - \alpha q + \alpha^2) + \alpha^2 (\alpha-2)}{2 q^2} e^{\frac{\alpha}{q}} \leq \\ \nonumber
& - \bar{\gamma} + \bar{\gamma} \left(1 - \frac{\alpha}{q} + \left(\displaystyle\frac{\alpha}{q} \right)^2 \right) e^{\frac{\alpha}{q}} - \bar{\gamma} \left(\frac{\alpha}{q} \right)^2 e^{\frac{\alpha}{q}} = \\
& \bar{\gamma} \left(-1 + e^{\frac{\alpha}{q}} \left( 1- \displaystyle\frac{\alpha}{q} \right) \right) \leq 0.
\end{eqnarray}
Here, the last inequality is due to the function
\begin{equation}\label{defin_mathcal_P}
\mathcal{P}(x) = (1+x) e^{-x}
\end{equation}
(with $x$ corresponding to $-\alpha/q$) satisfying $\mathcal{P}'(x) = -x e^{-x}$, which means that $\mathcal{P}(x)$ is increasing for $x<0$ and decreasing for $x>0$ and the maximum value is thus $\mathcal{P}(0)=1$. Note that the first inequality is strict provided $\alpha \neq 0$ (in which case $\partial \mathcal{L} / \partial q < 0 $ and we obtain uniqueness), and for $\alpha = 0$, we found explicit (unique) solutions in Section \ref{Section:Solution}.
\\
Thus we have found a region in the parameter space in which there exists a unique physically relevant solution $q(\alpha,\bar{\gamma})$ of \eqref{eq:mathcal L img}. 

\bibliographystyle{asmems4}

% Here's where you specify the bibliography database file.
% The full file name of the bibliography database for this
% article is asme2e.bib. The name for your database is up
% to you.
\bibliography{asme2e}

%%%%%%%%%%%%%%%%%%%%%%%%%%%%%%%%%%%%%%%%%%%%%%%%%%%%%%%%%%%%%%%%%%%%%%

\end{document}